\documentstyle[aps,epsfig]{revtex} 
\begin{document}
\title{Quasi-Normal Modes of General Relativistic Superfluid Neutron 
Stars}
\author{G. L. Comer}
\address{Department of Physics, Saint Louis University\\
P.O. Box 56907, St. Louis, MO 63156-0907, USA}
\author{David Langlois}
\address{D\'epartement d'Astrophysique Relativiste et de Cosmologie\\ 
CNRS, Observatoire de Paris, 92195 Meudon, France}
\author{Lap Ming Lin}
\address{Department of Physics, The Chinese University of Hong Kong\\
Hong Kong, China}

\date{\today}
	
\maketitle

\def\beq{\begin{equation}}
\def\eeq{\end{equation}}
\def\n{n}
\def\p{p}
\def\d{\delta}
\def\A{{\cal A}}
\def\B{{\cal B}}
\def\C{{\cal C}}
\def\Ha{{H_0}}
\def\Hb{{H_1}}
\def\H1{{H_1}}
\def\Vs{{V_\p}}
\def\Vn{{V_\n}}
\def\Wn{{W_\n}}
\def\Ws{{W_\p}}
\def\Xn{{X_\n}}
\def\Xs{{X_\p}}
\def\a00{{{\cal A}_0^0}}
\def\b00{{{\cal B}_0^0}}
\def\c00{{{\cal C}_0^0}}
\def\D00{{{\cal D}_0^0}}
\def\d{\delta}
\def\A{{\cal A}}
\def\B{{\cal B}}
\def\C{{\cal C}}
\def\M{{\cal M}}
\def\wh{\widehat}
\def\wa{\widetilde}
\def\ha{\overline}

\begin{abstract}
We develop a general formalism to treat, in general relativity, the 
linear oscillations of a two-fluid star about static (non-rotating) 
configurations.  Such a formalism is intended for neutron stars, 
whose matter content can be described, as a first approximation, by a 
two-fluid model: one fluid is the neutron superfluid, which is believed 
to exist in the core and inner crust of mature neutron stars; the other 
fluid is a conglomerate of all other constituents (crust nuclei, protons, 
electrons, etc...).  We obtain a system of equations which govern the 
perturbations both of the metric and of the matter variables, whatever 
the equation of state for the two fluids.  As a first application, we 
consider the simplified case of two non-interacting fluids, each with a 
polytropic equation of state.  We compute numerically the quasi-normal 
modes  (i.e. oscillations with purely outgoing gravitational radiation) 
of the corresponding system.  When the adiabatic indices of the two 
fluids are different, we observe a splitting for each frequency of the 
analogous single fluid spectrum.  The analysis also substantiates 
the claim that w-modes are largely due to spacetime oscillations.
\end{abstract}

\section{Introduction} \label{intro}

A formalism is presented here for describing the equilibrium 
configurations and quasi-normal modes of general relativistic neutron 
stars taking into account superfluidity.  There are both theoretical 
and observational reasons for believing that neutron stars have 
superfluid interiors.  On the theoretical side, nuclear physics 
calculations \cite{NP} indicate that the densities in neutron stars 
are favorable to the formation of ${}^1S_0$ neutron condensates in the 
neutron star crusts, with ${}^3P_2$ neutron condensates forming, as 
well as ${}^1S_0$ proton condensates (and perhaps even kaon and pion 
condensates), in the inner core.  On the observational side, there is 
the well-established glitch phenomenon, the best description of which 
is based on superfluidity and quantized vortices \cite{OA}.  Cooling 
rates for neutron stars are also best described by superfluidity 
\cite{Tsu}.

There are several distinct ways in which a superfluid can differ 
from an ordinary perfect fluid.  The most striking is that (pure) 
superfluids are  completely free of viscous effects and are locally 
irrotational.  The latter property is however compensated by the 
possibility for the superfluid to be threaded by quantized vortices, 
so that it can on macroscopic scales mimic the rotational behaviour 
of an ordinary fluid. 

In the context of neutron stars, one can distinguish three main fluid 
constituents: the neutrons, which can either belong to nuclei in the 
crust or be free and superfluid in the inner crust and the core; the 
protons, belonging only in nuclei in the crust and in a superconducting 
state in the core; and finally the electrons which behave everywhere in 
the star as a normal fluid.  As mentioned before, because of rotation, 
the neutron superfluid will contain vortices, whereas, because of the 
neutron star magnetic field, the proton superconductor is believed to 
contain magnetic fluxoids (at least if the core protons behave as a 
type II superconductor, which is the favored scenario \cite{uref1}). 
Finally, the whole picture is complicated by the entrainment effect 
(misleadingly called ``drag effect'' in the literature) between 
superconducting protons and superfluid neutrons, by which the momentum 
of one constitutent carries some mass current of the other constituent 
along with it (see e.g. \cite{uref1}).  The theoretical description of 
such mixtures, including the average effect of vortices (and fluxoids) 
was given by Lindblom and Mendell \cite{ML,M} in the Newtonian context, 
and more recently, by Carter and Langlois \cite{CL95,CL98} in the general 
relativistic context. 

In many astrophysical applications, the above picture can fortunately 
be simplified. Indeed, it has been shown that although the protons are 
superconducting, they are coupled to the ``normal'' electron fluid on 
a very short timescale \cite{uref2}. This implies, in practice, that 
the neutron star interior can be described, as a first step beyond the 
perfect fluid approximation, as a two-fluid model with a neutron 
superfluid component and a ``normal'' component containing everything 
else, i.e. the crust nuclei, the superconducting protons and the 
electrons.  Such a two-fluid description was presented by Langlois et 
al \cite{LSC98} in a general relativistic framework.  The corresponding 
formalism was based on a formally analogous but physically different 
model due to Carter (see, for instance, \cite{CL} and \cite{GCL}).  
The purpose of Carter's model was to extend to the general relativistic 
regime the non-relativistic two-fluid model of Landau (see \cite{P}), 
in which the ``normal'' fluid is constituted of excitations (phonons, 
rotons, etc.) of the condensate due to non-zero temperature.  In the 
case of neutron stars, the internal temperature is believed to decrease 
rather quickly after their  birth to temperatures well below the nuclear 
temperature scale (1 MeV), so that it will be justified, at least for 
the global structure of the star, to ignore the temperature effects. 

The effect of superfluidity on the oscillations of neutron stars 
has been investigated by Lindblom and Mendell \cite{LM1} in a 
{\it Newtonian} formalism.  Although analytical solutions for simple 
models revealed the existence of a new class of modes, which could be 
named {\it superfluid modes}, their numerical investigation \cite{LM1} 
led only to ordinary modes almost indistinguishable from those 
obtained by a single fluid treatment, with no sign of superfluid modes.  
However, a subsequent numerical treatment by Lee \cite{ulee}, still in 
the Newtonian regime, was successful in extracting the superfluid modes.   

The main objective of the present work is, in the same spirit as 
\cite{LM1,ulee}, to investigate the influence of superfluidity on the 
oscillations of neutron stars in the {\it general relativistic} context.  
The description of oscillations in general relativity is complicated 
essentially by  two factors: the first is that gravity is described by 
several components of the metric instead of the single gravitational 
potential of Newtonian gravity; the second is that oscillation of the 
star implies  emission of gravitational radiation, and therefore the 
oscillating solutions are damped (they are called {\it quasi-normal} 
modes instead of normal modes).

Due to the difficulties inherent to general relativity, we shall restrict 
our analysis to idealized models.  First, we consider only non-rotating 
neutron stars.  Second, for the numerical application, we consider an 
oversimplified situation in which the two fluids exist throughout the 
star (i.e. without taking into account the crust-core boundary, neutron 
drip) and are characterized by (independent) polytropic equations of 
state.  One of the fluids will be composed of neutrons, and the other 
will be a conglomerate made of protons and electrons, which we will call, 
loosely, ``proton'' fluid for simplicity in the rest of this paper.

The changes to the dynamics of the neutron star due to the existence of 
two fluids will be extracted to linear order via an analysis of the 
quasi-normal modes.  The linearized metric and matter variables will be 
decomposed into their respective ``even'' and ``odd'' parity components.  
This will generalize to the two-fluid case much of the previous work (e.g. 
\cite{DL,LD,chan,TC,PT,T1,T2,CT}) that has been done for the one-fluid 
case.  The present work can also be viewed as a first step in carrying 
over to the general relativistic regime the pioneering calculations of 
Lindblom and Mendell \cite{ML,M,LM1,LM2} for superfluid neutron stars 
in the non-relativistic regime.   

In Sec. \ref{sf} we describe the superfluid formalism.  In Sec. 
\ref{ec} we show how to construct equilibrium configurations.    
In Sec. \ref{lfe} we give the linearized field equations for purely 
radial and non-radial oscillations inside the star and for the vacuum 
outside the star.  The notation follows as closely as possible that 
used by \cite{DL,LD}.  In Sec. \ref{bc} the boundary conditions, 
numerical techniques, and quasi-normal mode (QNM) extraction are 
discussed.  The techniques used (which are much like those of 
\cite{DL,LD}) allow only the f-, p-, and w-modes to be reliably 
extracted.  In Sec. \ref{secAPP} the formalism is applied to the case 
of two relativistic polytropes.  It will be seen that the even-parity 
mode spectrum is different from that of neutrons alone, but only if the 
adiabatic index for the neutrons is different from that of the protons.  
In the first appendix we give the initial conditions that are used in 
the radial integration of the field equations.  The second appendix has 
a brief discussion of the analytic solution used to describe the vacuum 
spacetime exterior to the star.  ``MTW'' conventions \cite{MTW} apply 
throughout and the units are such that $G = c = 1$.  

\section{The two-fluid  formalism} \label{sf}

The central quantity of Carter's superfluid formalism is the 
so-called ``master'' function.  There are several choices for the 
master function.  Here it is taken to be the total thermodynamic 
energy density $- \Lambda$, which is a function of the three scalars 
$\n^2 = - \n_{\rho} \n^{\rho}$, $\p^2 = - \p_{\rho} \p^{\rho}$, and 
$x^2 = - \p_{\rho} \n^{\rho}$ that are formed from $\n^{\mu}$, the 
conserved neutron number density current, and $\p^{\mu}$, the 
conserved proton number density current.  The master function 
$\Lambda(\n^2,\p^2,x^2)$ encodes all information about the local 
thermodynamic state of the fluid and can also serve as a Lagrangian 
in an action principle for deriving the superfluid field equations 
(see, for instance, \cite{CL,GCL2}).  

A general variation (that keeps the spacetime metric fixed) of 
$\Lambda(\n^2,\p^2,x^2)$ with respect to the independent vectors 
$\n^{\mu}$ and $\p^{\mu}$ takes the form
\beq
     \d \Lambda = \mu_\rho \d \n^\rho + \chi_\rho \d \p^\rho \ ,
\eeq
where 
\beq
     \mu_{\mu} = \B \n_{\mu} + \A \p_{\mu} \quad , \quad
     \chi_{\mu} = \C \p_{\mu} + \A \n_{\mu} \ , \label{muchidef}
\eeq
and
\beq
   \A = - {\partial \Lambda \over \partial x^2} \qquad , \qquad \B = 
        - 2 {\partial \Lambda \over \partial \n^2} \qquad , \qquad
   \C = - 2 {\partial \Lambda \over \partial \p^2} \ . \label{coef1}
\eeq
The covectors $\mu_{\mu}$ and $\chi_{\mu}$ are dynamically, and 
thermodynamically, conjugate to $\n^{\mu}$ and $\p^{\mu}$, and their 
magnitudes are, respectively, the chemical potentials of the neutrons 
and the (conglomerate) protons.

The stress-energy tensor can be derived  (see, for instance, Ref. 
\cite{GCL,GCL2}) from $\Lambda$ and is found to be given by 
\beq
     T^{\mu}_{\nu} = \Psi \delta^{\mu}_{\nu} + \p^{\mu} \chi_{\nu} 
                   + \n^{\mu} \mu_{\nu} \ , \label{seten} 
\eeq
where $\Psi$ is the generalized pressure and is given by  
\beq
     \Psi = \Lambda - \n^{\rho} \mu_{\rho} - \p^{\rho} \chi_{\rho} \ .
\eeq
The equations of motion consist of two conservation equations,
\beq
     \nabla_{\mu} \n^{\mu} = 0 \qquad , \qquad \nabla_{\mu} \p^{\mu} = 0 
             \ , \label{coneq}
\eeq
and two Euler type equations, which can be conveniently written in 
the compact form 
\beq
     \n^{\mu} \nabla_{[\mu} \mu_{\nu]} = 0 \qquad , \qquad \p^{\mu} 
          \nabla_{[\mu} \chi_{\nu]} = 0 \ . \label{eueqn} 
\eeq
When all four are satisfied then it is automatically true that 
$\nabla_{\mu} T^{\mu}_{\nu} = 0$.  

Another property of this set of equations is the existence of two Kelvin 
theorems.  Defining the (antisymmetric) two-forms $w_{\mu \nu} = 
\nabla_{[\mu} \mu_{\nu]}$ and $\Omega_{\mu \nu} = \nabla_{[\mu} 
\chi_{\nu]}$, then the Lie derivative of $w_{\mu \nu}$ along 
$\n^{\mu}$ can be shown to be zero, and likewise for the Lie derivative 
of $\Omega_{\mu \nu}$ along $\p^{\mu}$.  Thus, if $w_{\mu \nu}$  
($\Omega_{\mu \nu}$) vanishes at some point on an integral curve of 
$\n^{\mu}$ ($\p^{\mu}$) then it must do so at all other points on the 
curve. 

As mentioned in the introduction, this same system of equations can 
describe the general relativistic analog of the Landau model for the 
non-relativistic superfluid \cite{P}.  In this case the vector 
$\n^{\mu}$ still represents a conserved particle number density current 
of the matter (which does not have to be neutrons) and $\mu_{\mu}$ is 
still its conjugate covector.  The proton number density current 
$\p^{\mu}$, however, gets replaced with $s^{\mu}$, which is the 
conserved entropy density current.  Likewise, $\chi_{\mu}$ is replaced 
by $\Theta_{\mu}$, the magnitude of which represents the local 
temperature of the fluid.

Since we will be mainly interested in the rest of this paper by linear 
variations of the matter and geometrical variables, let us introduce now 
the notation we will adopt in this respect. Following \cite{carter}, we
will write the variations of the momentum covectors 
$\mu_{\mu}$ and $\chi_{\mu}$ due to a generic variation of  
$\n^{\mu}$ and 
$\p^{\mu}$ and of the metric, in the following form,
\beq
    \d \mu_\rho = \A_\rho^{\ \sigma} \d \p_\sigma + \B_\rho^{\  \sigma}
                  \d \n_\sigma + \left(\d_g \A\right) \p_\rho + 
                  \left(\d_g \B\right) \n_\rho \ , \label{delmu}
\eeq
\beq
     \d \chi_\rho = \C_\rho^{\ \sigma} \d \p_\sigma + \A^\sigma_{\ 
                    \rho} \d \n_\sigma + \left(\d_g \C\right) \p_\rho + 
                    \left(\d_g \A\right) \n_\rho \ , \label{delchi}
\eeq
with
\begin{eqnarray}
\A_{\mu \nu} &=& \A g_{\mu \nu} - 2 {\partial \B \over \partial \p^2} 
              \n_\mu \p_\nu - 2{\partial \A \over \partial \n^2} \n_\mu 
              \n_\nu - 2 {\partial \A \over \partial \p^2} \p_\mu \p_\nu 
              - {\partial \A \over \partial x^2} \p_\mu \n_\nu \ , \cr
              && \cr
\B_{\mu \nu} &=& \B g_{\mu \nu} - 2 {\partial \B \over \partial \n^2} 
              \n_\mu \n_\nu - 4 {\partial \A \over \partial \n^2} 
              \p_{(\mu} \n_{\nu)} - {\partial \A \over \partial x^2} 
              \p_\mu \p_\nu \ , \cr
              && \cr
\C_{\mu \nu} &=& \C g_{\mu \nu} - 2 {\partial \C \over \partial \p^2} 
              \p_\mu \p_\nu - 4 {\partial \A \over \partial \p^2} 
              \p_{(\mu} \n_{\nu)} - {\partial \A \over \partial x^2} 
              \n_\mu \n_\nu \ ,
\end{eqnarray}
and the terms $\d_g \A$, $\d_g \B$ and $\d_g \C$ coming from the 
variation of the metric itself:
\beq
\d_g \A = \left[{\partial \A \over \partial \n^2} \n^\mu \n^\nu
          + {\partial \A \over \partial \p^2} \p^\mu \p^\nu + {\partial 
          \A \over \partial x^2} \n^\mu \p^\nu\right] \d g_{\mu\nu} \ 
\label{dgA}
\eeq
($\d_g \B$ and $\d_g \C$ being given by analogous formulas, where $\A$ 
is replaced by $\B$ and $\C$ respectively).  In order to solve the 
equations for the perturbations, it will be necessary to evaluate the 
coefficients $\A_{\mu \nu}$, $\B_{\mu \nu}$ and $\C_{\mu \nu}$ for the 
equilibrium configuration.

\section{The equilibrium configurations}\label{ec}

The background is spherically symmetric and static, so the metric can 
thus be written in the Schwarzschild form
\begin{equation}
  {\rm d}s^2 = - e^{\nu(r)} {\rm d}t^2 + e^{\lambda(r)} {\rm d}r^2 
               + r^2 \left({\rm d}\theta^2 + {\rm sin}^2\theta 
               {\rm d}\phi^2\right) \ . \label{bgmet}
\end{equation}
The non-trivial components of the Einstein tensor are
\begin{eqnarray}
   G^t_t &=& - \left({1 \over r} \lambda^{\prime} - {1 \over r^2}\right) 
             e^{-\lambda} - {1 \over r^2} \ , \label{einsteintt} \\
         &&  \cr
   G^r_r &=& {1 \over r} \left(\nu^{\prime} + {1 \over r}\right) 
             e^{- \lambda} - {1 \over r^2} \label{etcoms} \ , 
             \label{einsteinrr}\\
         &&  \cr
   G^\theta_\theta &=& G^\phi_\phi = \left({1 \over 2} \nu^{\prime 
             \prime} + {1 \over 4} \left(\nu^{\prime} - \lambda^{\prime}
             \right) \nu^{\prime} + {1 \over 2} \left(\nu^{\prime} - 
             \lambda^{\prime}\right) {1 \over r}\right)e^{- \lambda} \ .  
\end{eqnarray}
It is well known \cite{MTW} that the third of these components is 
{\it not} independent of the first two because of the Bianchi 
identities.

The description of the matter is much simplified here because of the 
time and spherical symmetry of the underlying spacetime; in particular, 
the two conserved currents $\n^{\mu}$ and $\p^{\mu}$ must be parallel 
with the timelike Killing vector $t^{\mu} = (1,0,0,0)$, i.e. that the 
independent vectors are of the form
\beq
    \p^{\mu} = \p u^{\mu} \ , \quad \n^{\mu} = \n u^{\mu} \ ,
\eeq
where $u^{\mu} = t^{\mu}/|{\bf t}|$.  Similarly, the dependent covectors 
are of the form 
\beq
    \chi_{\mu} = \chi(\n,\p) u_{\mu} \ , \quad \mu_{\mu} = \mu(\n,\p) 
                   u_{\mu} \ . \label{muchi}
\eeq
It should also be noted that $x^2 = \n \p$.  The stress-energy tensor 
also simplifies, with the non-zero components being
\beq
  T^t_t = \Lambda(\n,\p) \quad , \quad  T^r_r = T^{\theta}_{\theta} = 
          T^{\phi}_{\phi} = \Psi(\n,\p) = \Lambda + \mu \n + \chi \p 
          \ . \label{secoeffs} 
\eeq
The symmetry also implies some constants of the motion.  Indeed, from 
the vanishing of the Lie derivative of $\mu_{\mu}$ one finds
\beq
  2 t^\mu \nabla_{[\mu} \mu_{\nu]} + \nabla_\nu (t^\mu \mu_\mu)= 0 \ .
\eeq
The first term on the left-hand-side vanishes because of the equations 
of motion and one is left with 
\beq
     - \mu_{\infty} \equiv t^\mu \mu_\mu = \mu_t = const \ . \label{murs}
\eeq
The same procedure applies to $\chi_{\mu}$ and yields
\beq
     - \chi_{\infty} \equiv t^\mu \chi_\mu = \chi_t = const \ . 
                     \label{chirs}
\eeq
From Eqs. (\ref{muchidef}), (\ref{muchi}), (\ref{murs}), and 
(\ref{chirs}) one then finds
\beq
     \mu(\n,\p) = \mu_{\infty} e^{- \nu/2} = \B \n + \A \p \qquad 
                  , \qquad 
     \chi(\n,\p) = \chi_{\infty} e^{- \nu/2} = \A \n + \C \p \ . 
                 \label{flusol}
\eeq
It is simple to verify that $\n^{\mu}$, $\p^{\mu}$, $\mu_{\mu}$ and 
$\chi_{\mu}$ are now such that they automatically satisfy the Euler 
equations.  The two conservation equations for $\n^{\mu}$ and
$\p^{\mu}$ imply only that $\n$ and $\p$ are time-independent.  Taking
this with the spherical symmetry thus means $\n$ and $\p$ are functions
of $r$ only.  Thus, the fluid field equations have been completely 
exhausted.

On the other hand, the two independent Einstein equations---$G^t_t = 8 
\pi T^t_t$ and $G^r_r = 8 \pi T^r_r$---still remain to be solved.  
In the case of a one-fluid neutron star, these two equations are solved 
simultaneously with the TOV (Tolman-Oppenheimer-Volkoff) equation 
\cite{MTW}.  In our case of the two-fluid neutron star, we prefer to 
solve simultaneously a system of four equations that determine $\n(r)$, 
$\p(r)$, $\lambda(r)$, and $\nu(r)$.

The equations that govern the radial dependence of $\n$ and $\p$ are 
determined in the following way.  Starting from (\ref{flusol}), we 
obtain the following equations,
\beq
     \mu' = - \mu \nu'/2 \qquad , \qquad 
     \chi' = - \chi \nu'/2 \ . \label{muprim}
\eeq
One can now use the generic relations (\ref{delmu}) and (\ref{delchi}) 
to express $\mu'$ and $\chi'$ in terms of $\n'$ and $\p'$.  This 
requires us to evaluate the coefficients  $\A_{\mu \nu}$, etc., which 
will be needed anyway in the following for the equations governing the 
linear pertubations.  

For the equilibrium configuration, the coefficients involving spatial 
indices become
\beq
\A_{ij} = \A g_{ij}, \quad \B_{ij} = \B g_{ij}, \quad \C_{ij} = \C 
          g_{ij} \ ,
\eeq
and the mixed components, i.e. with one spatial and one time index, 
vanish altogether.  The most complicated coefficients are the 
time-time ones.  They are given explicitly by the following 
expressions
\begin{eqnarray}
\A_0^0 &=& \A + 2 {\partial \B \over \partial \p^2} \n \p + 2 
          {\partial \A \over \partial \n^2} \n^2 + 2 {\partial \A 
          \over \partial \p^2} \p^2 + {\partial \A \over \partial 
          x^2} \p \n \ , \cr
&& \cr
\B_0^0 &=& \B + 2 {\partial \B \over \partial \n^2} \n^2 + 4 
          {\partial \A \over \partial \n^2} \n \p + {\partial \A 
          \over \partial x^2} \p^2 , \cr
&& \cr
\C_0^0 &=& \C + 2 {\partial \C \over\partial \p^2} \p^2 + 4 {\partial 
           \A \over \partial \p^2} \n \p + {\partial A \over \partial 
           x^2} \n^2 \ . \label{coef2}
\end{eqnarray}
In the three coefficients above, after the partial derivatives are 
taken, then one sets $x^2 = \n \p$.   

The expressions (\ref{muprim}) can now be rewritten as differential 
equations that determine the radial profiles of $\n(r)$ and $\p(r)$: 
\beq
    \A^0_0 \p^{\prime} + \B^0_0 \n^{\prime} + {1 \over 2} (B \n + A \p) 
           \nu^{\prime} = 0 \quad , \quad
    \C^0_0 \p^{\prime} + \A^0_0 \n^{\prime} + {1 \over 2} (A \n + C \p) 
           \nu^{\prime} = 0 \ . \label{bgndfl}
\eeq
The functions $\nu(r)$, $\lambda(r)$, $\n(r)$, and $\p(r)$ can now be 
determined by solving these two equations in conjunction with the two 
independent Einstein equations, which can be written in the form
\beq
    \lambda^{\prime} = {1 - e^{\lambda} \over r} - 8 \pi r 
                         e^{\lambda} \Lambda(\n,\p) \quad , \quad
    \nu^{\prime} = - {1 - e^{\lambda} \over r} + 8 \pi r 
                         e^{\lambda} \Psi(\n,\p) \ . \label{bckgrnd}
\eeq  

This section is ended by recalling that a smooth joining of the interior 
spacetime to a Schwarzschild vacuum exterior at the surface of the star 
(the radial value of which we define to be $R$) implies two things: (i) 
the total mass $M$ of the system is given by
\beq
    M = - 4 \pi \int^{R}_0 {\rm d} r~r^2~\Lambda(r)
\eeq
and (ii) the total radial stress must vanish at the surface of the 
star, which implies $\Psi(R) = 0$.  In this work we will only look for 
background solutions that satisfy $\n(R) = 0$ and $\p(R) = 0$.  For the 
master function used later, these two conditions guarantee that $\Psi(R) 
= 0$ and also lead to $\Lambda(R) = 0$.  In view of (\ref{bckgrnd}), 
requiring a non-singular behaviour at the center of the star will impose 
that $\lambda(0) = 0$, and consequently that $\lambda^{\prime}(0)$ and 
$\nu^{\prime}(0)$ must also vanish. This in turn implies, in view of 
(\ref{bgndfl}), that $\p^{\prime}(0)$ and $\n^{\prime}(0)$ have to 
vanish as well.

\section{The linearized field equations} \label{lfe}

The metric and fluid perturbations will be decomposed on the basis of 
spherical harmonics $Y^m_l(\theta,\phi)$ (where $l$ is the angular 
momentum number and $m$ is its projection onto the $z$-axis) \cite{RW}.  
Because the background is spherically symmetric, there is no real loss 
of generality by restricting the study to the modes $m = 0$, i.e. by 
considering perturbations that do not depend on the $\phi$ coordinate.  
This means then that the basis on which the perturbations are actually 
decomposed is simply that of the Legendre polynomials $P_l(\theta) = 
Y_l^{m=0}(\theta,\phi)$.  

It will be seen to be convenient to distinguish between the so-called 
``even'' (i.e. $(-1)^l$-parity) components and the ``odd'' (i.e. 
$(-1)^{l + 1}$-parity) components.  For the even-parity modes a time 
dependence of the form
\begin{equation}
    \tilde{\cal O}(r,t) \equiv {\cal O}(r) e^{i \omega t} \ 
          \label{timedep}
\end{equation}
is assumed.  In general, $\omega$ is complex.  The same time-dependence 
for the odd-parity modes will not be imposed, because it will be shown 
later that there are no odd-parity pulsational modes for the fluid, but 
just differentially rotating ones (see also \cite{TC} for a discussion 
of the same result for one-fluid stars).  

Purely radial oscillations correspond to $l = 0$.  However, the 
system of equations written with general $l$ in mind must necessarily 
exclude the case of $l = 0$.  This makes it necessary, at the 
appropriate time, to consider the radial oscillations separately.

\subsection{The Linearized Metric and Matter Variables} \label{LM}

The form of the linearized Einstein Equations used is
\begin{equation}
    \delta R_{\mu \nu} =  8 \pi \delta \hat T_{\mu\nu} 
           \ , \label{leineq}
\end{equation}
where 
\begin{equation}
 \hat T_{\mu\nu}\equiv T_{\mu \nu} - {1 \over 2}
           T g_{\mu \nu}
\end{equation}
and $T\equiv g^{\mu\nu}T_{\mu\nu}$ is the trace of the 
stress-energy-momentum tensor.  The fluid equations to be linearized 
are the two conservation and two Euler equations, of which the latter 
two take the simple looking forms (for both parities)
\beq
    \partial_0 \d \mu_i = \partial_i \d \mu_0 \quad , \quad 
    \partial_0 \d \chi_i = \partial_i \d \chi_0 \ .
\eeq  

In the decomposition of the perturbations into even and odd-parity 
components, it can be seen that the ten linear perturbations of the 
metric will be divided into respectively seven and three components.  
Using then the four gauge transformations of general relativity, i.e. 
the coordinate transformations, three acting in the ``even'' subspace 
and one in the ``odd'' subspace, we can reduce the description of the 
metric perturbations to four ``even'' components and two ``odd'' 
components.  A convenient choice is the so-called  Regge-Wheeler gauge 
\cite{RW}, in which the even-parity components of the metric 
perturbations are
\begin{equation}
     \delta g_{\mu \nu} = - e^{i \omega t}
            \left[\matrix{e^{\nu(r)}r^l H_{0}(r)&i \omega r^{l+1} 
            H_{1}(r)&0&0 
            \cr i \omega r^{l+1} H_{1}(r)&e^{\lambda(r)} r^l 
            H_{2}(r)&0&0 \cr 
            0&0&r^{l+2} K(r)&0\cr 0&0&0&r^{l+2} {\rm sin}^2\theta K(r)}
            \right]  P_l(\theta) \ , \label{lemet}
\end{equation}
whereas the odd-parity components are
\beq
     \delta g_{\mu \nu} = \left[\matrix{0&0&0&
            h_{0}(r,t) {\rm sin}\theta{\partial \over \partial \theta}
             \cr 0&0&0& h_{\rm 1}(r,t) {\rm sin}\theta {\partial \over
            \partial \theta} \cr 0&0&0&0\cr h_{0}
            (r,t) {\rm sin}\theta{\partial \over \partial \theta}
            &h_{\rm 1}(r,t) {\rm sin}\theta{\partial \over \partial 
            \theta}&0&0}\right] P_l(\theta) \ . \ \label{lomet}
\eeq

To facilitate further manipulations of the linearized Euler equations, 
conservation equations, and $\delta \hat T_{\rho\sigma}$, $\n^{\mu}$ 
and $\p^{\mu}$ are rewritten as a product of a magnitude with a unit 
timelike vector so that
\beq
     \n_\rho = \n u_\rho \quad , \quad \p_\rho = \p v_\rho
\eeq
($u^{\rho} u_{\rho} = -1$ and $v^{\rho} v_{\rho} = -1$), where 
contrarily to the background case $u_\rho$ and $v_\rho$ are not aligned 
in general.  For our background, the spatial components of the 
perturbations of the momenta (recall Eqs. (\ref{delmu}) and 
(\ref{delchi})) are given by (true for both parities)
\beq
    \d\mu_i = \B \n\d u_i+\A \p\d v_i \quad ,\quad \d\chi_i = \C \p 
              \d v_i + \A \n \d u_i \ . \label{32}
\eeq
The time components are more difficult to obtain, but a straightforward 
calculation, using Eqs. (\ref{delmu}), (\ref{delchi}) and (\ref{dgA}) 
and inserting (\ref{lemet}) and ({\ref{lomet}), leads to 
\beq
\d\mu_0 = - e^{\nu/2} \left[\A_0^0\d \p + \B_0^0\d \n + {r^l \over 2} 
          \left(\B \n + \A \p\right)H_0 P_l e^{i \omega t}\right] \ ,
\eeq
and
\beq
\d\chi_0 = - e^{\nu/2} \left[\C_0^0\d \p + \A_0^0\d \n + {r^l \over 2} 
             \left(\A \n + \C \p\right)H_0 P_l e^{i \omega t}\right] \ ,
\eeq
for the even-parity components and $\d \mu_0 = \d \chi_0 = 0$ for the 
odd-parity components.  

The linearization and solving of the conservation equations, 
$\nabla_\mu \n^\mu = 0$ and $\nabla_\mu \p^\mu = 0$, are made easier 
(for even-parity modes) by introducing $\delta \xi_\n^i $ and $\delta 
\xi_\p^i $, which are the Lagrangian displacements, respectively, for 
the neutron and proton number density currents.  Now,
\beq
   \delta u^i = {{\rm d} \over {\rm d} \tau_\n} \delta \xi_\n^i \approx 
              e^{- \nu/2} {\partial \over \partial t} \delta \xi_\n^i 
              \quad , \quad
   \delta v^i = {{\rm d} \over {\rm d} \tau_\p} \delta \xi_\p^i \approx 
              e^{- \nu/2} {\partial \over \partial t} \delta \xi_\p^i \ ,
              \label{dxi} 
\eeq
where $\tau_\n$ ($\tau_\p$) is the proper time as measured by an 
observer whose four-velocity is $u^{\mu}$ ($v^{\mu}$).  A decomposition 
of the Lagrangian displacements into spherical harmonics yields
\beq
   \delta \xi_\n^r = e^{-\lambda/2}r^{l - 1}  \Wn(r) P_l 
                      e^{i \omega t} \quad , \quad
   \delta \xi_\n^{\theta} = - r^{l - 2} \Vn(r) 
                {\partial \over \partial \theta} P_l e^{i \omega t} 
                  \ , 
\eeq
and
\beq
   \delta \xi_\p^r = e^{-\lambda/2} r^{l - 1}  \Ws(r) P_l 
                      e^{i \omega t} \quad , \quad
   \delta \xi_\p^{\theta} = - r^{l - 2} \Vs(r) 
                {\partial \over \partial \theta} P_l e^{i \omega t} 
                  \ . 
\eeq

Insertion of the Lagrangian displacements into the conservation 
equations castes these equations into a form that can easily be 
integrated, the results being
\beq
  {\Delta n(r) \over n(r)} = - r^l \left(e^{-\lambda/2} 
            \left[{l + 1 \over r^2} \Wn + {1 \over r} \Wn^{\prime}
            \right] + {l (l + 1) \over r^2} V_n - {1 \over 2} H_2 - 
            K\right) \label{38}
\eeq
for the Lagrangian variation in the neutron number density, and
\beq
     {\Delta \p(r) \over \p(r)} = - r^l\left(e^{-\lambda/2} 
               \left[{l + 1 \over r^2} \Ws + {1 \over r} \Ws^{\prime}
               \right] + {l (l + 1) \over r^2} \Vs - {1 \over 2} H_2 - 
               K\right)\label{39}
\eeq
for the Lagrangian variation in the proton number density.  These are 
related to the Eulerian variations $\d \n(r)$ and $\d \p(r)$ via (see 
\cite{MTW}, pg. 691)
\begin{eqnarray}
   \Delta \n &=& \delta \n + \n^{\prime}  e^{-\lambda/2}
                r^{l - 1} \Wn \cr
            && \cr
   \Delta \p &=& \delta \p + \p^{\prime}  e^{-\lambda/2} 
                r^{l - 1} \Ws \ . \label{40}
\end{eqnarray}

For the odd-parity modes, $\delta \n(r) = 0$ and $\delta \p(r) = 0$, 
and the only non-zero components for $\d u_{\mu}$ and $\d v_{\mu}$ are
\beq
     \d u_{\phi} = e^{- \nu/2} \dot{U}_{\n}(r,t) 
                     {\rm sin} \theta {\partial P_l \over \partial 
                     \theta} \quad , \quad
     \d v_{\phi} = e^{- \nu/2} \dot{U}_{\p}(r,t) 
                     {\rm sin} \theta {\partial P_l \over \partial 
                     \theta} \ . 
\eeq
It can be shown that the conservation equations are automatically 
satisfied in this case.  

\subsection{Equations for Radial Oscillations} \label{RO}

When $l = 0$ more gauge freedom exists to eliminate some of the metric 
perturbations.  This freedom will be used to set $H_1 = 0$ and $K = 0$.  
Since the Legendre polynomial $P_0$ is just a constant, then all of 
the odd-parity perturbations and the Lagrangian displacements $\delta 
\xi^{\theta}_\n$ and $\delta \xi^{\theta}_\p$ vanish.  The only 
remaining perturbations are $H_0$, $H_2$, $\Wn$, and $\Ws$.  The 
remaining field equations consist of four linearized Einstein equations 
and two linearized Euler equations for the fluid.  

An independent set of equations contain two for the metric, which are
\begin{eqnarray}
   H^{\prime}_0 &=& 4 \pi r e^{\lambda} \left(\p^2 \C^0_0 + 2 \n \p 
       \A^0_0 + \n^2 \B^0_0 - 2 \Psi - {1 \over 4 \pi r^2}\right) H_2 - 
       \cr
        && \cr 
        &&{8 \pi e^{\lambda/2} \over r} \left(\left[\n \A^0_0 + \p 
        \C^0_0\right] \left[r \p \Ws\right]^{\prime} + \left[\p \A^0_0 + 
        \n \B^0_0\right] \left[r n \Wn\right]^{\prime}\right) \cr
       && \cr
   H_2 &=& 8 \pi e^{\lambda/2} \left(\chi \p \Ws + \mu n 
           \Wn\right) \ ,
\end{eqnarray}
and then two more for the fluid perturbations,
\begin{eqnarray}
   {\omega^2 \over r} e^{(\lambda - \nu)/2} \left(\A \p \Ws + \B \n 
       \Wn\right) &=& \left(- {e^{(\nu - \lambda)/2} \over r^2} 
       \left[\A^0_0 \left(r \p \Ws\right)^{\prime} + \B^0_0 \left(r \n 
       \Wn\right)^{\prime}\right] + \right. \cr
    && \cr
    && \left.{1 \over 2} e^{\nu/2} \left[\A^0_0 \p + \B^0_0 \n\right] 
       H_2\right)^{\prime} + {1 \over 2} \mu H^{\prime}_0 \ , \cr
    && \cr
    {\omega^2 \over r} e^{(\lambda - \nu)/2} \left(\C \p \Ws + \A \n 
       \Wn\right) &=& \left(- {e^{(\nu - \lambda)/2} \over r^2} 
       \left[\C^0_0 \left(r \p \Ws\right)^{\prime} + \A^0_0 \left(r \n 
       \Wn\right)^{\prime}\right] + \right. \cr
    && \cr
    && \left.{1 \over 2} e^{\nu/2} \left[\C^0_0 \p + \A^0_0 \n\right] 
       H_2\right)^{\prime} + {1 \over 2} \chi H^{\prime}_0 \ .
\end{eqnarray}
The two equations for the metric perturbations can be used in the two 
for the fluid perturbations to reduce the total system to two coupled 
second-order differential equations for $\Ws$ and $\Wn$.  This 
generalizes to our case the one-fluid results of Chandrasekhar 
\cite{chan}, who was able to go further and demonstrate that the final 
equation for the perturbations is of the Sturm-Liouville form.  Whether 
or not our system forms such a self-adjoint problem remains to be 
established. 

\subsection{Equations for Non-radial Oscillations} \label{NRO}

To obtain the equations for the even-parity oscillations, we have 
followed the formulation of Detweiler and Lindblom \cite{DL}, 
the main difference in our case arising from the existence of two 
fluids instead of one.  The procedure to obtain the equations is the 
following: the first step is to write explicitly the linearized Ricci 
tensor on the left-hand-side of (\ref{leineq}) in terms of the 
quantities introduced in (\ref{lemet}); the second step is to write 
the right-hand-side of (\ref{leineq}) in terms of the matter variables 
$\Ws$, $\Wn$, $\Vn$ and $\Vs$, as well as the metric variables, 
using (\ref{32})-(\ref{40}).  One then obtains six relations. In the 
case of a single fluid, there are enough equations.  However, in the 
case of two fluids, additional equations are necessary.  They will be 
given by the equations of motion for each fluid.  The conservation 
equations are automatically satisfied as a result of our choice of 
Lagrangian variables.  What remain are the Euler equations. This will 
provide us with four equations, corresponding to a radial component 
and an angular component for each fluid.  The whole system of 
equations is now redundant because of the Bianchi identities. 

After tedious calculations (performed by hand, and independently with 
the aid of Mathematica), we found that the Einstein and fluid 
equations yield a constraint equation
\begin{eqnarray}
 &&e^\lambda\left[{2-l-l^2\over r^2}- {3 \over r^2}\left(1-e^{- 
\lambda}\right)-8\pi \Psi\right]\Ha 
+\left[{2 \omega^2 \over e^{\nu}}-{l(l+1)\over 2}e^\lambda\left({1-e^{- \lambda} 
\over r^2} +8\pi \Psi\right)\right]\H1\cr
&&
+\left[-2 e^{\lambda-\nu}\omega^2+e^\lambda{l^2+l-2\over r^2}
+e^{2\lambda}\left({1-e^{- \lambda} \over r^2}+8\pi \Psi
\right)\left(1-{3\over 2}\left(1-e^{- \lambda}\right)- 4 \pi r^2 \Psi
\right)\right]K \cr
&& 
+16 \pi e^{\lambda-{\nu\over 2}}\left(\Xn + \Xs\right)=0 \ , 
\label{cnstrt}
\end{eqnarray}
having no radial derivatives, and the rest, which are first-order in the 
radial derivative, given by (using the notation $\D00=\b00\c00-\a00^2$)
\begin{eqnarray}
\H1^{\prime} &=& {e^\lambda\over r}\Ha 
+\left[{\lambda'-\nu'\over 2}-{l+1\over r}\right]
\H1
+{e^\lambda\over r} K - 16 \pi{e^\lambda\over r}\left(\mu \n \Vn + \chi
\p \Vs\right) \ , \label{h1prime} \\
&& \cr
K^{\prime} &=& {\Ha \over r} + {l (l+1) \over 2 r} \H1
+ \left[{\nu^{\prime} \over 2} - {l + 1 \over r}\right] K
-8 \pi {e^{\lambda/2} \over r}\left[\mu \n \Wn+\chi \p \Ws\right] \ ,
\\
&& \cr
\Wn^{\prime} &=& {e^{\lambda/2}r\over 2} \Ha
 + e^{\lambda/2} r K - e^{\lambda/2} {l (l+1)\over r} \Vn - 
 \left({l + 1 \over r} + {n^{\prime} \over \n}\right) \Wn + 
\cr
&&
+{\c00 \over \n^2 \D00}\left[e^{(\lambda-\nu)/2} r \Xn + n^{\prime} 
 \left(\b00 \n \Wn + \a00 \p \Ws\right)\right] - \cr
&&
{\a00 \over \n \p \D00}\left[e^{(\lambda-\nu)/2}r \Xs + \p^{\prime} 
 \left(\a00 \n \Wn + \c00 \p \Ws\right)\right] \ , \label{Wnprim} \\
&& \cr
\Ws^{\prime} &=& {e^{\lambda/2} r \over 2} \Ha
 + e^{\lambda/2} r K - e^{\lambda/2} {l (l + 1) \over r} \Vs - 
\left({l + 1 \over r} + {\p^{\prime} \over \p}\right) \Ws
\cr
&&
+{\b00 \over \p^2 \D00}\left[e^{(\lambda-\nu)/2} r \Xs + \p^{\prime} 
\left(\c00 \p \Ws+\a00 \n \Wn\right)\right] - \cr
&&
{\a00\over \n \p \D00}\left[e^{(\lambda-\nu)/2}r \Xn + \n^{\prime} 
 \left(\a00 \p \Ws+\b00 \n \Wn\right)\right] \ ,
\\
&& \cr
\Xn^{\prime} &=& - {l \over r} \Xn+{e^{\nu/2} \over 2} \left[\n \mu 
 \left({1 \over r}- \nu^{\prime}\right) - \n^{\prime} \left(\b00 \n 
 + \a00 \p\right)\right]\Ha \cr
&&
+ \mu \n \left[{e^{\nu/2} \over 4}{l (l + 1) \over r}
+ {\omega^2 \over 2} r e^{-\nu/2}\right] \H1 \cr
&&
+ e^{\nu/2}\left[\mu \n \left({\nu^{\prime} \over 4} - {1 \over 2 r}
 \right) - \left(\b00 \n + \a00 \p\right) \n^{\prime}\right]K
 + {l (l + 1) \over r^2} e^{\nu/2} \n^{\prime} \left(\b00 \n \Vn + 
 \a00 \p \Vs\right)
\cr
&&
-e^{(\lambda-\nu)/2} {\omega^2 \over r} \n \left(\B \n \Wn + \A \p 
 \Ws\right) - 4 \pi e^{(\lambda+\nu)/2} {\mu \n \over r} \left(\mu 
\n \Wn + \chi \p \Ws\right)
\cr
&&
+e^{-(\lambda-\nu)/2} \left[- {\n^{\prime} \over r} \left(\b00^{\prime} 
 \n \Wn + \a00^{\prime} \p \Ws\right)
+ \left({2 \n^{\prime} \over r^2} + {\lambda^{\prime} - \nu^{\prime} 
\over 2 r} \n^{\prime} - {\n^{\prime \prime} \over r}\right)
\right.\cr
&&
\left.\left(\b00 \n \Wn + \a00 \p \Ws\right)\right] \ , \\
&& \cr
\Xs^{\prime} &=& - {l \over r} \Xs + {e^{\nu/2} \over 2} \left[\p 
 \chi\left({1 \over r} - \nu^{\prime}\right)
 - \p^{\prime} \left(\c00 \p + \a00 \n\right)\right] \Ha \cr
&&
 + \chi \p \left[{e^{\nu/2} \over 4}{l (l + 1) \over r}
 + {\omega^2 \over 2} r e^{-\nu/2}\right] \H1 \cr
&&
 + e^{\nu/2} \left[\chi \p \left({\nu^{\prime} \over 4} - {1 \over 
 2r}\right) - \left(\c00 \p + \a00 \n\right) \p^{\prime}\right] K
 + {l (l + 1) \over r^2} e^{\nu/2} \p^{\prime} \left(\c00 \p \Vs + 
 \a00 \n \Vn\right)
\cr
&&
 - e^{(\lambda-\nu)/2} {\omega^2 \over r} \p \left(\C \p \Ws + \A \n 
 \Wn\right) - 4 \pi e^{(\lambda+\nu)/2} {\chi \p \over r} \left(\chi 
 \p \Ws + \mu \n \Wn\right)\cr
&&
 + e^{-(\lambda-\nu)/2} \left[-{\p^{\prime} \over r} \left(
 \c00^{\prime} \p \Ws + \a00^{\prime} \n \Wn\right)
 + \left({2 \p^{\prime} \over r^2} + {\lambda^{\prime}-\nu^{\prime}
 \over 2 r} \p^{\prime} - {\p^{\prime \prime} \over r}\right)\right.\cr
&&
\left.\left(\c00 \p \Ws + \a00 \n \Wn\right)\right] \ , \label{xpprime}
\end{eqnarray}
with 
\beq
\Xn \equiv \n\left[{e^{\nu/2}\over 2}\mu \Ha + e^{-\nu/2} \omega^2 
  \left(\B \n \Vn + \A \p \Vs\right)\right]
 - e^{(\nu-\lambda)/2} {\n^{\prime} \over r} \left(\b00 \n \Wn + \a00 
 \p \Ws\right) \ , \label{xn}
\eeq
\beq
\Xs \equiv \p \left[{e^{\nu/2} \over 2}\chi \Ha + e^{-\nu/2} \omega^2 
  \left(\C \p \Vs + \A \n \Vn\right)\right]
  - e^{(\nu-\lambda)/2} {\p^{\prime} \over r} \left(\c00 \p \Ws + 
  \a00 \n \Wn\right) \ . \label{xs}
\eeq
Note that $H_2$ has disappeared.  This is because the Einstein 
equations (for the present case of $l > 0$) impose $H_2 = H_0$, in 
contradistinction to the purely radial case.  

Let us emphasize that the form of our system of equations 
is quite analogous to that given in \cite{DL} for the one-fluid case.  
The main difference in our system is the doubling of the number of 
equations corresponding to the fluid degrees of freedom (i.e. the $W$'s 
and $V$'s).  Note also that an explicit coupling between the two fluids 
is manifest in the latter equations when the coefficients $\A_0^0$ or 
$\A$ are non-zero. If they are zero, this means the two fluids are 
independent but they remain  coupled indirectly through the gravitational 
field, i.e. the equations governing the degrees of freedom of the fluids 
include a dependence on the metric perturbations.  

For odd-parity perturbations the Euler equations yield
\beq
     \A \p \ddot{U}_{\p} + \B \n \ddot{U}_{\n} = 0 \qquad , \qquad
     \C \p \ddot{U}_{\p} + \A \n \ddot{U}_{\n} = 0 \ , \label{opeueqs}
\eeq
whereas the non-trivial linearized Einstein equations are
\begin{eqnarray}
     \left({1 \over e^{\lambda}} \left[{\nu^{\prime} - \lambda^{\prime} 
     \over 2 r} + {1 \over r^2}\right] - {l [l + 1] \over 2 r^2} \right) 
     h_1 - {1 \over 2 e^{\nu}} \ddot{h}_1 + {1 \over 2 e^{\nu}} 
     \left(\dot{h}^{\prime}_0 - {2 \over r} \dot{h}_0\right) 
     &=& \cr
     && \cr
     4 \pi \left(\Psi + \Lambda\right) h_1 \ , \cr
     && \cr
     {1 \over e^{\nu}} \dot{h}_0 - {1 \over e^{\lambda}} \left(
     h^{\prime}_1 + {\nu^{\prime} - \lambda^{\prime} \over 2} h_1\right) 
      &=& 0 \ , \cr
     && \cr
     {1 \over e^{\lambda}} \left(h^{\prime \prime}_0 - {\nu^{\prime} + 
     \lambda^{\prime} \over 2} h^{\prime}_0 + {2 \nu^{\prime} \over r} 
     h_0 - \left[\dot{h}^{\prime}_1 + \left({2 \over r} - {\nu^{\prime} 
     + \lambda^{\prime} \over 2}\right) \dot{h}_1\right]\right) - {l (l 
     + 1) \over r^2} h_0 
     &=& \cr
     && \cr
     16 \pi \left(\chi \p \dot{U}_{\p} + \mu n \dot{U}_{\n}\right) + 8 \pi 
     \left(\Psi + \Lambda\right) h_0 \ . &&
\end{eqnarray}
Recall that odd-parity modes automatically satisfy the conservation 
equations.  The Euler equations thus imply that $\ddot{U}_{\n}$ and 
$\ddot{U}_{\p}$ both vanish, which means at this order the fluid only 
differentially rotates, and does not pulsate.  This is similar to what 
happens in the ordinary one-fluid case \cite{TC}.  If pulsations 
are considered, then the linearized Euler equations imply that the 
fluid perturbations vanish, and then the gravitational degrees of 
freedom are decoupled from those of the fluid.  

\subsection{The Linearized Equations Outside the Star} \label{VE}

Outside the star, the fluid variables are all zero.  The problem is 
then reduced to the metric perturbations alone, which can be 
conveniently obtained from the so-called Zerilli function $Z$ 
\cite{ZER}, which satisfies
\beq
     {{\rm d}^2 Z \over {\rm d}{r^*}^2} = \left(V_z(r) - \omega^2 
                  \right) Z \ ,
\eeq
where $r^* = r + 2 M {\rm ln}\left(r/2M - 1\right)$, $M$ is the stellar 
mass, $n = (l - 1)(l + 2)/2$, and the effective potential is
\beq
     V_Z(r) = {1 - 2M/r \over r^3 (n r + 3 M)^2} \left(2 n^2 (n + 1) r^3 
              + 6 n^2 M r^2 + 18 n M^2 r + 18 M^3\right) \ .
\eeq
As $r \to r^* \to \infty$, then the asymptotic form of $Z$ is
\beq
    Z \sim A_{in} e^{i \omega r^*} + A_{out} e^{- i \omega r^*} \ . 
      \label{AF}
\eeq

The relations \cite{Fack,LD} that determine $H_0$ and $K$ from $Z$ are
\beq
    Z\left(r^*\right) = {a(r) H_0(r) + (b(r) - k(r)) K(r) \over h(r) - 
                        k(r) g(r)} \label{Zer}
\eeq
and
\beq
    {{\rm d}Z\left(r^*\right) \over {\rm d}r^*} = K(r) - g(r) 
                        Z\left(r^*\right) \ , \label{dZer}
\eeq
where
\begin{eqnarray}
   a(r) &=& - {n r + 3 M \over \omega^2 r^2 - (n + 1) M/r} \ , \cr
    && \cr
   b(r) &=& {n r (r - 2M) - \omega^2 r^4 + M (r - 3 M) \over (r - 2 M) 
            \left(\omega^2 r^2 - (n + 1) M/r\right)} \ , \cr
    && \cr
   g(r) &=& {{n (n + 1) r^2 + 3 n M r + 6 M^2} \over r^2 (n r + 3 M)} 
       \ , \cr
    && \cr
   h(r) &=& {- n r^2 + 3 n M r + 3 M^2 \over (r - 2 M) (n r + 3 M)} \ , 
            \cr
    && \cr
   k(r) &=& - {r^2 \over r - 2 M} \ .
\end{eqnarray}
(It should be mentioned here that $b(r)$ suffers from a typographical 
error in \cite{LD}, but was subsequently corrected in \cite{DL}.)  The 
remaining metric perturbation $H_1$ can be obtained from the vacuum 
form of the constraint equation (\ref{cnstrt}), i.e.
\begin{eqnarray}
      &&\left[3 M + {1 \over 2}(l + 2)(l - 1) r\right] \Ha = - \left[
        {l (l + 1) \over 2} M - {\omega}^2 r^3 e^{- (\nu + \lambda)}
        \right] \Hb \cr
      &&+ \left[{1 \over 2}(l + 2)(l - 1) r - {\omega}^2 r^3 e^{-\nu}
        - {1 \over r} e^{\lambda} M (3 M - r)\right] K \ .
\end{eqnarray}

\subsection{The r\^ole of vortices} \label{secIF}

A superfluid differs from an ordinary fluid in that it is locally 
irrotational.  It can nevertheless closely mimic rotation via the 
formation of an array of quantized vortices that threads the fluid.  
Such an array has not been explicity included here.  Since our initial 
configuration is not rotating, it contains no vortices.  When there are 
no vortices, the superfluid is said to be in a Landau state \cite{TT}.  
A further restriction must be imposed to obtain a strict Landau state.  
This restriction is that the vorticity $w_{\mu \nu} = 
\nabla_{[\mu} \mu_{\nu]}$ vanishes.  This is already automatically 
satisfied by the background configuration. 

If one wished to impose the irrotationality condition on the 
perturbations as well, then it can be shown that this would restrict 
the set of allowed perturbations by imposing the constraint $H_1(r) = 0$ 
in the case of even-parity and $\B \n \dot{U}_\n + \A \p \dot{U}_\p = 0$ 
in the case of odd-parity perturbations.  However, a small perturbation 
at the scale of the whole star will be felt at a mesoscopic level as a 
huge perturbation in the sense that many vortices will form spontaneously.  
This can be illustrated for example by a rough estimate of the critical 
angular velocity for the formation of one vortex.  For a Newtonian 
laboratory superfluid, this value is given by (see e.g. \cite{TT}) 
\beq
\Omega_{c1}\simeq {\hbar\over 2 m_n R^2}\ln\left({R\over a}\right) \ ,
\eeq
where $a$ is the size of the vortex core.  For a neutron star, 
$a \sim 10^{-12}$ cm and $R \sim 10$ km, so that 
$\Omega_{c1} \sim 10^{-14}$ ${\rm rad/s}$, which is extremely 
small.  This means that, even for small perturbations of the star, one 
expects a huge number of vortices so that the superfluid will mimic an 
ordinary perfect fluid on macroscopic scales.  It thus makes perfect 
sense to treat the superfluid as we have done here. 

\section{Boundary Conditions, Numerical Techniques, and QNM Extraction} 
\label{bc}

\subsection{Boundary Conditions at $r = 0$ and $r = R$} \label{secBC}

In Sec. \ref{secAPP} numerical solutions to the even-parity system of 
equations (\ref{cnstrt})-(\ref{xpprime}) are constructed.  An essential 
element of this construction is to know the behaviour of the fields near 
$r = 0$, because of a general problem of spherical coordinates that 
one cannot start a numerical integration precisely at $r = 0$, but 
only from a nearby point.  As outlined by \cite{DL}, the values of the 
fields near $r = 0$ can be obtained by expanding each field in a Taylor 
series, and then using the field equations to determine the coefficients 
of these expansions.  The results, which are presented in the first 
appendix, show that one need only specify the set of values 
$\{K(0),\Wn(0),\Ws(0)\}$, for instance, and then the remainder, $H_1(0)$, 
$\Xn(0)$, and $\Xs(0)$, are determined.  All of the second derivatives, 
$H_1^{\prime \prime}(0)$, $K^{\prime \prime}(0)$, etc, are likewise 
determined by $\{K(0),\Wn(0),\Ws(0)\}$.

At the surface, the story is somewhat different.  In the case of a 
single fluid star, the boundary condition is that the pressure should 
vanish at the {\it physical} surface of the star.  Since the fluid has 
moved with respect to its equilibrium value, the physical surface is 
displaced with respect to the background star surface and the 
appropriate condition is thus that the Lagrangian variation of the 
pressure vanishes at the surface of the star.  The same reasoning cannot 
be applied here because we have two Lagrangian variations to consider, 
one along the neutrons and another along the protons.  However, in our 
particular example where the two fluids are independent, one can 
generalize easily the single fluid result because the pressure is 
separable, in the sense  that it can be decomposed unambiguously into 
two pressures corresponding respectively to each fluid, $\Psi = \Psi_n 
+ \Psi_\p$ (with $\Psi_n = \Lambda_n + \mu n$, etc.).  The physically 
meaningful boundary condition is then to impose that the Lagrangian 
variation of each pressure should vanish at the star surface.  Let us 
express this condition in terms of our variables.  The Lagrangian 
variation of $\Psi_n$ for instance will be given by
\beq 
\Delta \Psi_n =n\Delta\mu=n\B_0^0\Delta n \ .
\eeq
Substituting the expression (\ref{38}) and using $H_2 = H_0$ as well as
(\ref{Wnprim}), one gets the simple relation
(like in the one-fluid case of course)
\beq
\Delta \Psi_n =-e^{-\nu/2}r^l \Xn \ ,
\eeq
and the corresponding result for $\Delta \Psi_\p$.  Our boundary 
conditions will thus be $\Xn(R) = 0$ and $\Xs(R) = 0$.  

\subsection{Numerical integration for quasi-normal modes} 

One finds in the numerical integration the closest parallels between 
the formalism presented here and what has been done for the perfect 
fluid with one constituent \cite{DL,LD}.  Given a background 
configuration, the three main components of the numerical procedure 
are (i) integrate inside the star, (ii) integrate outside the star, 
and (iii) match the interior and exterior solutions at the surface 
of the star, and find those $\omega$ (corresponding to quasi-normal 
modes) for which there are only outgoing gravitational waves at 
spatial infinity.  We use the dgear subroutine available from the 
IMSL library to numerically integrate the background equations, and 
the linearized equations inside and outside of the star.

The integration inside the star has itself two separate parts.  The 
perturbation equations (\ref{h1prime})-(\ref{xpprime}) in the interior 
can be written as the matrix equation
\beq
    {{\rm d}{\bf Y} \over {\rm d}r} = {\bf Q} \cdot {\bf Y} \ , 
             \label{ME}
\eeq
where
\beq
     {\bf Y} = \{H_{1},K,W_{\n},W_{\p},X_{\n},X_{\p}\}
\eeq
is an abstract six-dimensional vector field.  The $6 \times 6$ matrix 
${\bf Q}$ depends on $l$, $\omega$, the background fields, and the 
various coefficients $\A$, $\a00$, etc.  As mentioned earlier, the 
results presented in the first appendix indicate that only the set 
$\{K(0),\Wn(0),\Ws(0)\}$ needs to be specified and then the remaining 
quantities $H_1(0)$, $\Xn(0)$, and $\Xs(0)$, and all of the second 
derivatives at $r = 0$, are determined.  At the surface of the star, 
there are only the conditions that $X_{\n}(R) = X_{\p}(R) = 0$, so that 
one must specify the set of values $\{H_1(R),K(R),\Wn(R),\Ws(R)\}$.  

For a given background, the first part to the interior integration 
consists of choosing three arbitrary values of $\{K(0),\Wn(0),\Ws(0)\}$, 
and then starting the integration of Eq. (\ref{ME}) at small $r_{0}$, 
where 
\beq
    {\bf Y}(r_{0}) = {\bf Y}(0) + {1 \over 2} 
                     {\bf Y}^{\prime \prime}(0) r_0^2 \ . 
\eeq
(Both vectors ${\bf Y}(0)$ and ${\bf Y}^{\prime \prime}(0)$ can be 
constructed from the results presented in the first appendix.)  The 
integration is terminated at $r = R/2$.  This process is repeated two 
more times, each with a different choice for $\{K(0),\Wn(0),\Ws(0)\}$, 
to get three linearly independent solutions ${\bf Y}_{1}$, 
${\bf Y}_{2}$, and ${\bf Y}_{3}$, and hence a general solution in the 
domain $0 \leq r \leq R/2$ of the form
\beq
    {\bf Y}(r) = \sum^3_{i = 1} c_{i} {\bf Y}_{i}(r) ,
\eeq
where the $c_{i}$ ($i = 1,2,3$) are constants to be determined.

The second part of the interior integration starts from the surface of 
the star.  There are four linearly independent solutions ${\bf Y}(r)$ 
that can be constructed.  These are built by choosing four arbitrary 
sets of values $\{H_1(R),K(R),\Wn(R),\Ws(R)\}$ and then integrating 
backward from $R$ to $R/2$ four times to generate a general solution 
in the domain $R/2 \leq r \leq R$ of the form
\beq
    {\bf Y}(r) = \sum^7_{i = 4} c_{i} {\bf Y}_{i}(r) \ ,
\eeq 
where the $c_{i}$ ($i = 4,5,6,7$) are constants to be determined.

At $r = R/2$, the two general solutions must be equal:  
\beq
    \sum^3_{i = 1} c_{i} {\bf Y}_{i}(R/2) = \sum^7_{i = 4} c_{i} 
           {\bf Y}_{i}(R/2) \ . \label{EQ}
\eeq
Each value of ${\bf Y}_i(R/2)$ ($i = 1,...,7$) is of course known.  
Hence, picking arbitrarily one of the $c_i$ and giving it some value, 
then Eq. (\ref{EQ}) represents six equations that can be used to 
determine the remaining six $c_i$.  Having obtained the $c_{i}$, 
${\bf Y}(r)$ is thus known throughout the star.  

Next, the metric perturbations outside the star must be determined. A
priori, since all perturbations are known throughout the star, the
boundary values $H_0(R)$ and $K(R)$ obtained from the interior solutions
are sufficient as initial data to determine all the metric perturbations
outside the star, embodied in the particular Zerilli function $Z_{reg}$ 
which satisfies, according to (\ref{Zer}) and (\ref{dZer}),
\begin{eqnarray}
     Z_{reg} &=& {a(R) H_0(R) + (b(R) - k(R)) K(R) \over h(R) -
                        k(R) g(R)} \ , \cr
             && \cr
     \left.{{\rm d}Z_{reg} \over {\rm d}r^*}\right|_{r = R} &=& K(R) -
             g(R) Z_{reg} \ . \label{zmatch}
\end{eqnarray}
However such a procedure will give in general a mixture of ingoing and
outgoing gravitational radiation, whereas we wish to determine the
specific values of $\omega$ for which there is only outgoing
gravitational radiation.  For the conventions used here, this implies
that $A_{in} = 0$ in the asymptotic form of the Zerilli function (cf. 
Eq. (\ref{AF})).  It has long been known, however, that this is 
problematic numerically, because stable stars undergoing quasi-normal 
mode oscillations must have (for the conventions used here) 
${\rm Im}(\omega) > 0$ and this implies that the exponential multiplying 
$A_{out}$ in Eq. (\ref{AF}) diverges as $r^* \to \infty$.  We use two 
different techniques to circumvent this difficulty.  One will enable an 
accurate determination of the so-called w-modes, and the other an 
approximation of the f- and p-modes.

To determine the w-modes we use the Leaver series \cite{L1,L2} in a
manner like that of \cite{YTL,LLSTY,CWY}.  It is, albeit indirectly,
an analytic representation for $Z_{out}$, the Zerilli function that
describes an outgoing wave at spatial infinity.  (The details of this
series are given in the second appendix.) Like $Z_{reg}$, $Z_{out}$
depends only on $l$, $\omega$ and $M$, so that for given $l$ and $M$ the 
only free parameter is $\omega$.  The special values of $\omega$ that 
correspond to quasi-normal modes are extracted by solving for the (in 
general, complex) roots of $f(\omega) = 0$, where
\beq
    f(\omega) \equiv \left.{1 \over Z_{reg}} {{\rm d} Z_{reg} \over
              {\rm d}r^*}\right|_{r = R} - \left.{1 \over Z_{out}}
              {{\rm d} Z_{out} \over {\rm d}r^*}\right|_{r = R} \ .
\eeq
We use Muller's method \cite{NR} to determine the roots.

In principal, the procedure just outlined should succeed in determining 
all the quasi-normal modes.  In practice, it only works well for the 
w-modes.  The difficulty lies in the fact that the ratio ${\rm Im}
(\omega)/{\rm Re}(\omega)$ of the fluid modes are typically so small 
($<10^{-4}$; see also \cite{T1}) that the iteration of Muller's method 
is not possible.  (In practice, one can only obtain the f-modes by 
using very accurate initial guesses.)  To determine the f- and p-modes 
a direct integration is done, with $Z_{reg}$ and 
${\rm d}Z_{reg}/{\rm d}r^*$ (cf. Eq. (\ref{zmatch})) as initial 
conditions, to obtain the Zerilli function.  The problem of the 
divergence of $Z$ is overcome by looking only for resonant solutions 
that have real $\omega$.  The true quasi-normal modes are not produced, 
but accurate numbers for the real part of the frequencies should be 
obtained since ${\rm Im}(\omega)/{\rm Re}(\omega) < 10^{-4}$.  The 
radial profiles of all the fields should thus also be representative of 
the true quasi-normal modes.  In general, direct integration will yield 
a $Z$ that corresponds to both incoming and outgoing waves at spatial 
infinity.  The resonant frequencies that correspond to an outgoing wave 
alone are extracted via a ``graphical'' technique that maps out 
${\rm log}\left|A_{in}\right|$ vs $\omega$.  Those values of $\omega$ 
that cause deep minima to occur in ${\rm log}\left|A_{in}\right|$ 
correspond to the resonant frequencies.

\subsection{Convergence Tests and Accuracy}

Convergence tests have been performed to determine the reliability and 
accuracy of the numerical results.  The tests discussed here are for 
the model three presented in Table I, which is for a star with neutrons 
and (conglomerate) protons,  each fluid behaving as a relativistic 
polytrope (cf. Eq. (\ref{EOS})).  Also, we have considered only $l = 2$ 
modes. 

The accuracy of the results for the w-modes is determined by 
calculating the changes in the w-mode frequencies for different step 
sizes.  The integrations have been performed using (scaled) step sizes 
of $\Delta r = 10^{-5}$ (which corresponds to $\Delta r/R \sim 
10^{-4}$) and $\Delta r = 5 \times 10^{-6}$.  The fractional change 
in the real part of the frequency is defined by
\beq
     {{\rm Re}(\Delta \omega) \over {\rm Re}(\omega)} \equiv 
              {{\rm Re}(\omega_2) - {\rm Re}(\omega_1) \over 
              {\rm Re}(\omega_1)}\ ,
\eeq
where $\omega_1$ is the frequency calculated using the first step size 
and $\omega_2$ is for the second.  A similar definition is employed
for the imaginary part of $\omega$.  The results for the first six 
w-modes are summarized in Table II.

Another test has been done by changing the matching point inside the 
star (cf. the numerical integration discussion of the previous 
subsection).  Two different matching points, $R/2$ and $R/4$, have been 
used.  It is found that the fractional change in the frequencies is 
$\Delta \omega/\omega \sim 10^{-4}-10^{-5}$, which is comparable to 
the changes of Table II.  One can be confident, then, that the w-mode 
frequencies produced by our numerical scheme have at least three 
significant figures accuracy. 

The accuracy of the f- and p-modes is determined by varying the 
truncation point $r_{trunc}$ which is the final grid point of the direct 
integration that determines the Zerilli function.  As discussed earlier, 
the resonant frequencies are located by finding the deep minima of the 
incoming wave amplitude $A_{in}$, which is obtained from 
\beq
     A_{in} = \left({e^{- i \omega r^*} \over 2}\left.\left(Z + {1 \over 
              i \omega} {{\rm d}Z \over {\rm d}r^*}\right)\right)
              \right|_{r = r_{trunc}} \ .
\eeq 
The tests reveal that good convergence is achieved when $r_{trunc} > 50 
R$.  It was also found that the resonant frequency corresponding to the 
first deep minimum can be used as a good initial guess in the Leaver's 
series method, and agrees very well with the real part of the f-mode 
obtained from the Leaver's series ($\omega M = 0.130 + 6.52 \times 
10^{-5} i$).  A value of $r_{trunc} = 300 R$ will be used in the next 
section.  

\section{Application: Two Relativistic Polytropes} \label{secAPP}

We will now apply our formalism to the concrete example where each fluid 
behaves as a relativistic polytrope.  Let us take a master function of 
the form
\beq
    \Lambda(\n^2,\p^2) = - m_\n \n - \sigma_{\n} \n^{\beta_{\n}}
                              - m_\n \p - \sigma_{\p} \p^{\beta_{\p}}
                              \ , \label{EOS}
\eeq
where we have assumed for simplicity the same mass $m_\n$ (i.e. the 
neutron mass) for both fluid particles.  This master function is not 
only independent of $x^2$ it is also separable.  An immediate consequence 
is that $\A = \a00 = 0$.  The other relevant coefficients are given by
\beq
 \B = {\mu\over \n} = {m_\n \over \n} + \sigma_\n \beta_\n
      \n^{\beta_\n - 2} \quad , \quad \b00 = \sigma_\n \beta_\n
      \left(\beta_\n - 1\right) \n^{\beta_\n - 2} \ ,
\eeq
and  
\beq
 \C = {\chi \over \p} = {m_\n \over \p} + \sigma_\p \beta_\p
      \p^{\beta_\p - 2} \quad , \quad \c00 = \sigma_\p \beta_\p
      \left(\beta_\p - 1\right)\p^{\beta_\p - 2} \ .
\eeq
The two fluids are also assumed to be in chemical equilibrium for the 
background, which means that the condition $\mu = \chi$ is imposed, and 
therefore $\n$ and $\p$ are linked by an algebraic relation valid 
throughout the star.  Thus, only one central particle density needs to 
be specified in order to determine the star model.  It should be noted 
that chemical equilibrium is entirely consistent with Eqs. (\ref{flusol}) 
and (\ref{muprim}) (or Eq. (\ref{bgndfl})), since these equations 
themselves imply that if $\mu = \chi$ at one point in the star then 
$\mu = \chi$ at all points.   

The discussion will be limited to two particular choices for the free
parameters $\sigma_{\n}$, $\sigma_{\p}$, $\beta_{\n}$ and $\beta_{\p}$,
which are the two different sets listed for models one through four in
Table I.  In Fig. 1 we show the dependence of the mass $M$ on the total
central density, i.e. $\n_c = \n_0 + \p_0$, using the ``$\sigma$''
and ``$\beta$'' values of models one through four in Table I.  Fig. 2
contains the radial profiles of $n$ and $p$ for models one (left graph)
and two (right graph).  One can see in Fig. 1 the existence of maxima 
at around $\n_c = 2.5~fm^{- 3}$ for both sets of ``$\sigma$'' and 
``$\beta$'' values.  The branches of these curves to the left of the 
maxima represent those configurations that are stable to radial 
perturbations.  Thus, the two sets of values for $\n_0$ and $\p_0$ 
chosen for models one and two (to be used in the quasi-normal mode 
analysis below) correspond to stable configurations; models three and 
four are unstable.  

\subsection{The w-modes}

The w-modes were first discovered by Kokkotas and Schutz \cite{KS} for 
one-fluid stars.  They are due primarily to the oscillations of 
spacetime itself, coupling only weakly to the fluid of the star.  (See, 
for instance, Andersson et al \cite{Aetal}, who extracted w-modes 
using an Inverse Cowling Approximation, wherein all the fluid motion is 
frozen out.)  Another characteristic property of w-modes is that they 
are strongly damped (${\rm Im}(\omega) \sim {\rm Re}(\omega)$).  Lein 
et al \cite{Lein} have found a branch of strongly damped w-modes, which 
are denoted ${\rm w_{II}}$, that are similar to the quasi-normal modes 
of black holes. 
 
In Fig. 3 we plot, for $l = 2$, several w-mode frequencies for models 
one and two, and in Table III we list the explicit values for the real 
and imaginary parts for the first six modes of Fig. 3 for both models.  
The ${\rm w_{II}}$-modes are the two whose frequencies satisfy 
${\rm Im}(\omega) > {\rm Re}(\omega)$.  There are no qualitative 
differences for the w-modes between our models one and two, nor are 
there such differences between both models and what one finds for a 
one-fluid system.  Figs. 4 and 5 are radial profiles (for the 
${\rm w_1}$- through ${\rm w_4}$-modes listed in Table III) of $W_\n$ 
(which also describe $W_\p$ since it is indistinguishable from $W_\n$ 
for these cases) for models one and two, respectively.  Again, there 
are no qualitative differences between the two models for $W_\n$ 
(or $W_\p$).  

It is also important to point out that Figs. 4 and 5 demonstrate that 
the neutrons and the protons move in ``lock-step'' with each other for 
each w-mode frequency and for both models (the density variations show 
the same).  This can be considered to be further evidence for the idea 
that w-modes are largely spacetime oscillations, because below it will 
be seen that the obviously fluid oscillation f- and p-modes allow for 
counter-moving motion between the neutrons and the protons.

\subsection{The f- and p-modes}

The f-mode, or fundamental mode, may be regarded as the ``simplest'' 
non-radial mode.  It has no radial nodes and it usually reaches a 
maximum value near the surface of the star \cite{Metal}.  Physically 
one may understand this mode as a surface wave that maintains the 
total volume of the system.  The p-modes, or pressure-modes, have as 
their restoring force pressure and they behave like sound waves.  
Typically there is an infinite number of p-mode frequencies.  The 
radial profile of the first p-mode, which will be denoted 
p${}_{\rm 1}$, will contain one radial node; the second mode, denoted 
by p${}_{\rm 2}$, will have two; and so on.  Unlike the w-modes, 
there are qualitative differences between models one and two for the 
f- and p-modes.

Fig. 6 is a graph of ${\rm log}|A_{in}|$ vs ${\rm Re}(\omega M)$, 
with $l = 2$, for model one and also for a star composed solely of 
neutrons behaving as a relativistic polytrope (with the same neutron 
parameter values as model one).  Reading it from left to right, the 
first deep minimum corresponds to the f-mode, with the next 
corresponding to the first p-mode, and so on.  One should notice that 
the deep minima for the one-fluid star are virtually indistinguishable 
from the star with two fluids.  

Fig. 7 is a similar graph, but for model two of Table I.  The obvious 
difference with Fig. 6 is the doubling of the f-mode, and (near) 
doubling of the p-modes.  For a more precise numerical comparison, 
Table IV lists the real parts of the f- and p${}_1$-mode frequencies for 
model one, and their ``doubled'' counterparts from model two, which are 
denoted f${}_{\rm o}$ (``o'' for ordinary), f${}_{\rm s}$ (``s'' for 
superfluid), p${}_{\rm 1o}$, and p${}_{\rm 1s}$.  

The new, i.e. superfluid, modes represent the case where the neutrons 
and protons no longer move in ``lock-step'' with each other.  This is 
illustrated most dramatically, perhaps, by Figs. 8, 9, and 10.  Fig. 
8 contains the radial profiles of $\Delta \n/\n$ (or $\Delta \p/\p$, 
since they are identical in this case) for the ($l = 2$) f- and 
p${}_{\rm 1}$-modes of model one; Fig. 9 gives both $\Delta \n/\n$ and 
$\Delta \p/\p$ for the ($l = 2$) f${}_{\rm o}$- and f${}_{\rm s}$-modes 
of model two; and, Fig. 10 gives $\Delta \n/\n$ and $\Delta \p/\p$ 
for the ($l = 2$) p${}_{\rm 1o}$- and p${}_{\rm 1s}$-modes of model 
two.  For the ordinary modes the two density variations increase and 
decrease on the same intervals, but for each of the superfluid modes the 
two density variations are clearly out of phase, i.e. the neutron number 
density increases when the protron density decreases, and vice versa.  

Figs. 11, 12, and 13 contain (for $l = 2$) radial profiles of $W_\n$ and 
$W_\p$ for models one and two.  Fig. 11 is for model one, and not 
surprising there is no difference between the radial flow of the neutrons 
and the protons.  However, Figs. 12 and 13 are for model two, and clearly 
evident is counter-motion between the neutrons and the protons for the 
superfluid modes, i.e. the neutrons are flowing in when the protons are 
flowing out, and vice versa.    

Finally, with Figs. 14-17, one can contrast the effects of the 
superfluid with the ordinary fluid modes on the spacetime perturbations 
$H_1$ and $K$.  We have also considered higher $l$ values, and have 
found superfluid modes.  Table IV lists the higher $l$ frequencies 
for the f- and ${\rm p_1}$-modes for both models one and two.  

\subsection{w-, f-, and p-modes for Radially Unstable Configurations}

The w-, f-, and p-modes have been investigated for configurations that 
are unstable to purely radial oscillations, that is, they are located 
to the right of the maxima of Fig. 1.  These are models three and four 
of Table I.  The general features of the quasi-normal modes discussed 
above for models one and two carry over to models three and four.  Thus 
even stars on the unstable branch contain the counter-moving modes. 

\section{Concluding Remarks}

The main purpose of this work was to build a formalism appropriate to 
study the quasi-normal modes of general relativistic neutron stars with 
superfluidity.  We have described the matter content of a neutron star 
in terms of simply two components, one corresponding  to a neutron 
superfluid and another corresponding to a proton-electron conglomerate 
fluid.  This approximation makes sense since protons and electrons can 
be seen as locked together, from their electromagnetic interaction, on 
bulk motion scales so that they move in ``lock-step'' with each other 
as the star oscillates.  We have ignored effects due to the crust, as 
well as electromagnetic effects that would lead to, for instance, the 
magnetic fields that neutron stars are known to possess.

We have performed a numerical investigation based on the system of 
equations that we obtained for linear perturbations both of the fluids 
and of the metric.  We have found numerically general relativistic 
superfluid modes, i.e. a new set of modes characterizing the dual 
nature of the star matter content which doubles the number of degrees 
of freedom in the matter.  In the Newtonian regime, these modes had 
been revealed analytically for a simplified model by Lindblom and 
Mendell \cite{LM1,LM2,ML,M} and found numerically by Lee \cite{ulee}.  

A distinguishing characteristic of these new modes is that they are 
counter-moving.  Also, when the adiabatic indices of the two constituents 
are equal then there is a degeneracy between these new modes and their 
analogs in one-fluid stars.  Lindblom and Mendell argued that the 
superfluid mode frequencies should be larger than the ordinary mode 
frequencies.  Our work verifies this conclusion for the general 
relativistic regime: Table IV clearly shows that ${\rm f_o} < 
{\rm f_s}$ and ${\rm p_{1o}} < {\rm p_{1s}}$.  We have also verified 
the inequalities for the higher node p-modes.

Finally, our results also tend to support the claim that w-modes are due 
mostly to spacetime oscillations, since there is very little qualitative 
difference between our models one and two.  The obviously fluid f- and 
p-modes, on the other hand, were clearly distinguished between models 
one and two, exhibiting the counter-moving modes.

\acknowledgements

G.L.C. gratefully acknowledges financial support from the Graduate 
School of Saint Louis University, the Programme on Initiatives on 
Numerical Relativity and General Relativistic Astrophysics of the 
Chinese University of Hong Kong, and the Department of Relativistic 
Astrophysics and Cosmology of the Observatory of Paris.  D.L. would 
like to acknowledge Saint Louis University and Washington University 
for their hospitality and financial support.  L.M.L. gratefully 
acknowledges financial support from Washington University.  We also 
thank Nils Andersson, Ming Chung Chu, Pui Tang Leung, Wai-Mo Suen, 
and Ching Wa Yip for useful discussions.  The research is supported 
in part by NSF Grant PHY 96-00507, NASA Grant NCCS 5-153, and the 
Hong Kong Grant Council Grant CUHK 4189/97p.

\section{Appendix I: radial integration initial conditions}

The background quantities will be expanded near $r = 0$ in the form 
\beq
q(r) = q_0 + {1 \over 2} q_2 r^2 + {1 \over 4} q_4 r^4 + {\cal O}(r^6).
\eeq
The field equations imply that the first and third derivatives of each 
field must vanish at $r = 0$.  We shall use an overbar and an overhat 
to designate the zeroth order and the second order respectively, so 
that, for instance,
\beq
\ha{\n} = \n_0 \qquad , \qquad \wh{\n} = {1 \over 2} \n_2 \qquad , 
       \qquad \ha{\mu \n} = \mu_0 \n_0 \qquad ,
\qquad \wh{\mu \n} = {1 \over 2} \left(\mu_0 \n_2 + \mu_2 \n_0\right).
\eeq
By comparison with a Taylor series expansion, one recognizes that 
\beq
q_0 = q(0), \quad q_2 = q^{\prime \prime}(0), \quad q_4 = {1 \over 6} 
      q^{(iv)}(0) .
\eeq

Now, the lowest order of the perturbation equations yield 
the following three constraints (where it is useful to know that 
$H_0(0) = K(0)$, $\Vn(0) = - \Wn(0)/l$, and $\Vs(0) = - \Ws(0)/l$):
\begin{eqnarray}
\Xn(0) &=& {e^{\nu_0/2} \over 2} \ha{\mu \n} K(0)
 - \left(e^{\nu_0/2} \n_2 \ha{\b00 \n} + {\omega^2 \over l} 
 e^{-\nu_0/2} \ha{\B \n^2}\right) \Wn(0)\cr
&&
 - \left(e^{\nu_0/2} \n_2 \ha{\a00 \p} + {\omega^2 \over l} 
 e^{-\nu_0/2} \ha{\A \n \p}\right) \Ws(0) \ , \\ \label{cnsrnt1}
&& \cr
\Xs(0) &=& {e^{\nu_0/2} \over 2} \ha{\chi \p} K(0)
 - \left(e^{\nu_0/2} \p_2 \ha{\c00 \p} + {\omega^2 \over l} 
 e^{-\nu_0/2} \ha{\C \p^2}\right) \Ws(0)\cr
&&
 - \left(e^{\nu_0/2} \p_2 \ha{\a00 \n} + {\omega^2\over l} 
 e^{-\nu_0/2} \ha{\A \n \p}\right) \Wn(0) \ , \\ \label{cnsrnt2} 
&& \cr
\H1(0) &=& {2 \over l + 1} K(0) + {16 \pi \over l (l + 1)} 
  \left[\ha{\mu \n} \Wn(0) + \ha{\chi \p}\Ws(0)\right] \ , 
  \label{cnsrnt3} 
\end{eqnarray}

For the next order, it is useful to recognize that the background 
Einstein equations imply
\beq
\lambda^{\prime \prime}(0) = - {16 \pi \over 3} \Lambda_0, \quad 
\nu^{\prime \prime}(0) = {8 \pi \over 3} \left(3 \Psi_0 - \Lambda_0
\right) \ .
\eeq
It is also useful to know that
\beq
    H_0^{\prime \prime}(0) = K^{\prime \prime}(0) + Q_0 
\eeq
and
\beq 
\Vn^{\prime \prime}(0) = Q_\n - {l + 3 \over l (l + 1)} 
\Wn^{\prime \prime}(0) \quad , \quad
\Vs^{\prime \prime}(0) = Q_\p - {l + 3 \over l (l + 1)} 
\Ws^{\prime \prime}(0) \ , 
\eeq
where
\begin{eqnarray}
Q_0 &\equiv& {4 \over (l + 2)(l - 1)} \left[8 \pi e^{-\nu_0/2}
            \left(\Xn(0) + \Xs(0)\right) - \left(- {8 \pi\over 3} 
            \Lambda_0 + \omega^2 e^{-\nu_0}\right) K(0)\right.
\cr
    && \left.-\left({2 \pi l (l + 1) \over 3}(3 \Psi_0 - \Lambda_0)
       - \omega^2 e^{-\nu_0}\right) \H1(0)\right], \\
&& \cr
Q_\n &\equiv& {2 \over l (l + 1)} \left\{{3 \over 2} K(0) -
\left({4 \pi (l + 1) \over 3} \Lambda_0 + {\n_2 \over \n_0}\right)
\Wn(0)\right.\cr
&&\left.+ \ha{\left({\c00 \over \n^2 \D00}\right)} \left[e^{-\nu_0/2} 
\Xn(0) + \n_2 \left(\ha{\b00 \n} \Wn(0) + \ha{\a00 \p} \Ws(0)\right)
\right]\right.\cr
&&\left. - \ha{\left({\a00 \over \n \p \D00}\right)}
\left[e^{-\nu_0/2} \Xs(0) + \p_2 \left(\ha{\a00 \n} \Wn(0) + 
\ha{\c00 \p} \Ws(0)\right)\right]\right\} \ ,\\
&& \cr
Q_\p &\equiv& {2 \over l(l + 1)} \left\{{3 \over 2} K(0) -
\left({4 \pi (l + 1) \over 3} \Lambda_0 + {\p_2\over \p_0}\right) 
\Ws(0)\right.\cr
&&\left.+ \ha{\left({\b00 \over \p^2 \D00}\right)}
\left[e^{-\nu_0/2} \Xs(0) + \p_2\left(\ha{\c00 \p}\Ws(0) + 
\ha{\a00 \n} \Wn(0)\right)\right]\right.\cr
&&\left. -
\ha{\left({\a00\over \n \p \D00}\right)} \left[e^{-\nu_0/2} \Xn(0) + 
\n_2 \left(\ha{\a00 \p} \Ws(0) + \ha{\b00 \n} \Wn(0)\right)\right]
\right\} \ .
\end{eqnarray}
The next order of the perturbation equations then yield the following:
\begin{eqnarray}
&&{e^{-\nu_0/2} \over 2} \Xn^{\prime \prime}(0) - {1 \over 4} 
  \ha{\mu \n} K^{\prime \prime}(0) + {1 \over 2} \left[\n_2 \ha{\b00 \n} 
  + e^{-\nu_0} \omega^2{l + 3 \over l (l + 1)} \ha{\B \n^2}\right] 
  \Wn^{\prime \prime}(0)\cr
&& 
  + {1 \over 2} \left[\n_2 \ha{\a00 \p} + e^{-\nu_0} \omega^2 {l + 3 
  \over l (l + 1)} \ha{\A \n \p}\right] \Ws^{\prime \prime}(0) = 
  {e^{-\nu_0/2} \over 4} \nu_2 \Xn(0) + {1 \over 2} \wh{\mu \n} K(0) + 
  {1 \over 4}\ha{\mu \n} Q_0 \cr
&&
  + {e^{-\nu_0} \over 2} \omega^2 \left(\ha{\B \n^2} Q_\n
  + \ha{\A \p \n} Q_\p\right) - \left[\left(\n_4 + {4 \pi \Lambda_0 
  \over 3} \n_2\right) \ha{\b00 \n} + \n_2 \wh{\b00 \n} + {\omega^2 
  \over l e^{\nu_0}}\left(\wh{\B \n^2} - \right.\right. \cr
&&\left.\left.{\nu^{\prime \prime}(0) \over 2} \ha{\B \n^2}\right)\right]
  \Wn(0) - \left[\left(\n_4 + {4 \pi \Lambda_0 \over 3} \n_2\right)
  \ha{\a00 \p} + \n_2 \wh{\a00 \p} + {\omega^2 \over l e^{\nu_0}} 
  \left(\wh{\A \n \p} - \right.\right.\cr
&&\left.\left.{\nu^{\prime \prime}(0) \over 2} \ha{\A \p \n}\right)
  \right] \Ws(0) \ ,
\end{eqnarray}
\begin{eqnarray}
&&{e^{-\nu_0/2} \over 2} \Xs^{\prime \prime}(0) - {1 \over 4} 
  \ha{\chi \p} K^{\prime \prime}(0) + {1 \over 2} \left[\p_2 \ha{\c00 \p} 
  + e^{-\nu_0} \omega^2{l + 3 \over l (l + 1)} \ha{\C \p^2}\right] 
  \Ws^{\prime \prime}(0)\cr
&&
  + {1 \over 2} \left[\p_2 \ha{\a00 \n} + e^{-\nu_0} \omega^2 {l + 3 
  \over l (l + 1)} \ha{\A \n \p}\right] \Wn^{\prime \prime}(0) = 
  {e^{-\nu_0/2} \over 4} \nu_2 \Xs(0) + {1 \over 2} \wh{\chi \p} K(0) 
  + {1 \over 4}\ha{\chi \p} Q_0\cr
&&
  + {e^{-\nu_0} \over 2} \omega^2 \left(\ha{\C \p^2}Q_\p 
  + \ha{\A \p \n}Q_\n\right) - \left[\left(\p_4 + {4 \pi \Lambda_0 \over 
  3} \p_2\right) \ha{\c00 \p} + \p_2 \wh{\c00 \p} + {\omega^2 \over l 
  e^{\nu_0}} \left(\wh{\C \p^2} - \right.\right.\cr
&&\left.\left. {\nu^{\prime \prime}(0) \over 2} \ha{\C \p^2}\right)\right]
  \Ws(0) - \left[\left(\p_4 + {4 \pi \Lambda_0\over 3} \p_2\right)
  \ha{\a00 \n} + \p_2 \wh{\a00 \n} + {\omega^2 \over l e^{\nu_0}} 
  \left(\wh{\A \n \p} - \right.\right.\cr
&&\left.\left.{\nu^{\prime \prime}(0) \over 2} \ha{\A \n \p}\right)\right]
  \Wn(0) \ ,
\end{eqnarray}
\begin{eqnarray}
&&
{l + 2 \over 2} K^{\prime \prime}(0) - {l (l + 1) \over 4} 
   \H1^{\prime \prime}(0)+ 4 \pi \left[\ha{\mu \n} \Wn^{\prime \prime}(0) 
   + \ha{\chi \p} \Ws^{\prime \prime}(0)\right] = {1 \over 2} Q_0 \cr
&&
   + {1 \over 2} \nu_2 K(0) - 8 \pi \left[\left(\wh{\mu \n} - {4 \pi 
   \Lambda_0 \over 3}\ha{\mu \n}\right) \Wn(0) + \left(\wh{\chi \p} - {4 
   \pi \Lambda_0 \over 3}\ha{\chi \p}\right) \Ws(0)\right],
\end{eqnarray}
\begin{eqnarray}
&&
{l + 3 \over 2} \H1^{\prime \prime}(0) - K^{\prime \prime}(0) - {8 \pi 
   (l + 3) \over l (l + 1)} \left(\ha{\mu \n} \Wn^{\prime \prime}(0) + 
   \ha{\chi \p}\Ws^{\prime \prime}(0)\right) = - \left({8 \pi (l + 2) 
   \over 3} \Lambda_0 + {\nu_2 \over 2}\right) \times \cr
&& \H1(0)  + {16 \pi \over l} \left[\wh{\mu \n} \Wn(0) + \wh{\chi \p} 
   \Ws(0)\right] - 8 \pi \left(\ha{\mu \n} Q_\n + \ha{\chi \p} Q_\p\right) 
   + {1 \over 2} Q_0 \ ,
\end{eqnarray}
\begin{eqnarray}
&&{l + 2 \over 2} \Xn^{\prime \prime}(0) - {l (l + 1) \over 8} 
   e^{\nu_0/2} \ha{\mu \n} \H1^{\prime \prime}(0) + e^{\nu_0/2} 
   \left[{l + 2 \over 2} \n_2 \ha{\b00 \n} + 2 \pi (\ha{\mu \n})^2
   + {1 \over 2} e^{-\nu_0} \omega^2 \ha{\B \n^2}\right] \times \cr
&& \Wn^{\prime \prime}(0) + e^{\nu_0/2}\left[{l + 2 \over 2} \n_2 
   \ha{\a00 \n} + 2 \pi \ha{\mu \n \chi \p} + {1 \over 2}
    e^{-\nu_0} \omega^2 \ha{\A \p \n}\right]\Ws^{\prime \prime}(0) \cr
&& = \left({\wh{\mu \n} \over \ha{\mu \n}} + {1 \over 4} \nu_2\right) l 
   \Xn(0) - e^{\nu_0/2} \left[{1 \over 4}\ha{\mu \n} \nu_2 + {3 \over 2} 
   \n_2\ha{\left(\b00 \n + \a00 \p\right)}\right] K(0) \cr
&&+ {1 \over 2} \omega^2 e^{-\nu_0/2} \ha{\mu \n} \H1(0) + {1 \over 4} 
  e^{\nu_0/2}\ha{\mu n}Q_0+{l (l + 1) \over 2}e^{\nu_0/2}
  \n_2 \left(\ha{\b00 \n} Q_\n + \ha{\a00 \p} Q_\p\right) \cr
&&+ \left\{e^{\nu_0/2} \left[\left({\wh{\mu \n} \over \ha{\mu \n}}l 
  \n_2 - (l + 2) \n_4 - {1 \over 2} \n_2 \nu_2 - {4 \pi \over 3} 
  \Lambda_0 \n_2\right) \ha{\b00 \n} - l \n_2 \wh{\b00 \n} - 4 \pi 
  \ha{\mu \n} \wh{\mu \n} + \right.\right. \cr
&&\left.\left.{1 \over 3}\left(4 \pi \ha{\mu \n}\right)^2 \Lambda_0
  - \n_0 \n_2 \left(\b00\right)_2\right] + \omega^2 e^{-\nu_0/2}
  \left[\left({\wh{\mu \n}\over \ha{\mu \n}} + {1 \over 2}
  \nu_2 + {4 \pi \over 3} \Lambda_0\right) \ha{\B \n^2} - \wh{\B
  \n^2}\right]\right\} \Wn(0)
\cr
&&+ \left\{e^{\nu_0/2} \left[\left({\wh{\mu \n} \over \ha{\mu \n}}l 
  \n_2 - (l + 2) \n_4 - {1 \over 2} \n_2 \nu_2 - {4 \pi \over 3} 
  \Lambda_0 \n_2\right) \ha{\a00 \p} - l \n_2 \wh{\a00 \p} - 4 \pi 
  \ha{\mu \n} \wh{\chi \p} +\right.\right.\cr
&&\left.\left. {(4 \pi)^2 \over 3} \ha{\mu \n \chi \p}
  \Lambda_0 - \p_0 \n_2 \left(\a00\right)_2 \right] + \omega^2 
  e^{-\nu_0/2} \left[\left({\wh{\mu \n} \over \ha{\mu \n}} + {1 
  \over2} \nu_2 + {4 \pi \over 3} \Lambda_0\right) \ha{\A \p \n} - 
  \wh{\A \p \n}\right]\right\} \Ws(0) \ ,
\end{eqnarray}
\begin{eqnarray}
&&{l + 2 \over 2} \Xs^{\prime \prime}(0) - {l (l + 1) \over 8} 
   e^{\nu_0/2}\ha{\chi \p} \H1^{\prime \prime}(0) + e^{\nu_0/2} 
   \left[{l + 2 \over 2} \p_2 \ha{\c00 \p} + 2 \pi (\ha{\chi \p})^2
   + {1 \over 2} e^{-\nu_0} \omega^2 \ha{\C \p^2}\right] \times \cr
&& \Ws^{\prime \prime}(0) + e^{\nu_0/2} \left[{l + 2 \over 2} \p_2 
   \ha{\a00 \p} + 2 \pi \ha{\mu \n \chi \p} + {1 \over 2} e^{-\nu_0} 
   \omega^2 \ha{\A \p \n}\right] \Wn^{\prime \prime}(0) \cr
&&= \left({\wh{\chi \p} \over \ha{\chi \p}} + {1 \over 4} \nu_2\right) 
   l \Xs(0) - e^{\nu_0/2} \left[{1 \over 4} \ha{\chi \p} \nu_2 + {3 
   \over 2} \p_2 \ha{\left(\c00 \p + \a00 \n\right)}\right] K(0) \cr
&&+ {1 \over 2} \omega^2 e^{-\nu_0/2} \ha{\chi \p} \H1(0) + {1 \over 
  4} e^{\nu_0/2} \ha{\chi \p} Q_0 + {l (l + 1) \over 2} e^{\nu_0/2}
  \p_2 \left(\ha{\c00 \p} Q_\p + \ha{\a00 \n} Q_\n\right) \cr
&&+ \left\{e^{\nu_0/2} \left[\left({\wh{\chi \p} \over \ha{\chi \p}} l 
  \p_2 - (l + 2) \p_4 - {1 \over 2} \p_2 \nu_2 - {4 \pi \over 3} 
  \Lambda_0 \p_2\right) \ha{\c00 \p} - l \p_2 \wh{\c00 \p} - 4 \pi 
  \ha{\chi \p} \wh{\chi \p} + \right.\right.\cr
&&\left.\left.{1 \over 3} \left(4 \pi \ha{\chi \p} \right)^2 \Lambda_0
  - \p_0 \p_2 \left(\c00\right)_2\right] + \omega^2 e^{-\nu_0/2}
  \left[\left({\wh{\chi \p} \over \ha{\chi \p}} + {1\over 2}
  \nu_2 + {4 \pi \over 3} \Lambda_0\right) \ha{\C \p^2}-\wh{\C
  \p^2}\right]\right\} \Ws(0)
\cr
&&+ \left\{e^{\nu_0/2} \left[\left({\wh{\chi \p} \over \ha{\chi \p}} l 
  \p_2 - (l + 2) \p_4 - {1 \over 2} \p_2 \nu_2 - {4 \pi \over 3} 
  \Lambda_0 \p_2\right) \ha{\a00 \n} - l \p_2 \wh{\a00 \n} - 4 \pi 
  \ha{\chi \p} \wh{\mu \n} + \right.\right. \cr
&&\left.\left.{(4 \pi)^2 \over 3} \ha{\mu \n \chi \p} \Lambda_0 - \n_0 
  \p_2 \left(\a00\right)_2\right] + \omega^2 e^{-\nu_0/2} \left[
  \left({\wh{\chi \p} \over \ha{\chi \p}} + {1 \over 2} \nu_2 + {4 \pi 
  \over 3}\Lambda_0\right) \ha{\A \p \n} - \wh{\A \p \n}\right]\right\}
  \Wn(0) \ .
\end{eqnarray}

\section{Appendix II: The Leaver Series}

Leaver's series \cite{L1} is an analytic solution of the outgoing wave 
of the Regge-Wheeler equation \cite{RW}
\beq
    {{\rm d}^2\phi_{RW} \over {{\rm d}r^*}^2} = \left[V_{RW}(r) - 
               \omega^2\right] \phi_{RW} \ ,
\eeq
where
\beq
    V_{RW}(r) = \left(1 - {2M \over r}\right) \left[{l (l + 1) \over r^2} 
                 + (1 - s^2) {2M\over r^3}
\right]\ .
\eeq
We will consider the case $s=2$, which is for gravitational waves; $s=0$ 
and $s=1$ correspond to the massless scalar field and electromagnetic 
waves, respectively.  

Letting $g$ represent the outgoing wave solution, then its analytic 
expression is
\beq
     g(\bar{\omega},\bar{r}) = \bar{r}^{1 + s} (\bar{r} - 1)^{\rho} 
           e^{-\rho \bar{r}} \sum_{n=0}^{\infty} a_n (2 \rho + 1)_n 
           U(s + 1 + 2 \rho + n,2 s + 1,2 \rho \bar{r})\ ,
\eeq
where $\bar{r} = r/2M$ and $\rho = - i \bar{\omega} = - i 2 M \omega$,  
$U$ is the irregular confluent hypergeometric function, and 
\beq
     (z)_n \equiv {\Gamma(z + n) \over \Gamma(z)}
\eeq
is the Pochhammer's symbol.  The coefficients $a_n$ are determined by 
the recurrence relation
\begin{eqnarray}
    \alpha_n a_{n + 1} + \beta_n a_n + \gamma_n a_{n - 1} &=& 0 \ \ \ \ 
             n=1,2... \cr
        && \cr
    a_n &=& 0 \ \ \ \ n < 0 \ ,
\end{eqnarray}
where $a_0$ is arbitrary and
\begin{eqnarray}
     \alpha_n &=& (n + 1) (n + 2 \rho + 1) \ , \cr
             && \cr 
     \beta_n &=& - \left[2 n^2 + (8\rho + 2) n + 8 {\rho^2} + 4 \rho + 
                 l (l + 1) + (1 - s^2)\right]\ , \cr
             && \cr
     \gamma_n &=& n^2 + 4 \rho n + 4 {\rho}^2 - s^2 \ .
\end{eqnarray}
It must be remarked that, in the derivation of $g(\bar{\omega},\bar{r})$, 
the time dependence is assumed to be $e^{-i\omega t}$, which has a minus 
sign different from our convention $e^{i\omega t}$.

The even-parity perturbation outside the star is described by the Zerilli 
function, while the Leaver's series gives the analytic expression for the
outgoing wave solution of Regge-Wheeler equation.  Thus, we have to 
employ transformations between the Regge-Wheeler and Zerilli functions in 
order to apply Leaver's series.  The transformations are \cite{chan1}
\beq
    \left[{\mu}^2 \left({\mu}^2 + 2\right) + 12 i \omega M\right] Z = 
    \left[{\mu}^2 \left({\mu}^2 + 2\right) + {72 M^2 (r - 2 M) \over r^2 
    \left({\mu}^2 r + 6 M\right)}\right] \phi_{RW} + 12 M {{\rm d} 
    \phi_{RW} \over {\rm d}r^*}\ , 
\eeq
\beq
    \left[{\mu}^2 \left({\mu}^2 + 2\right) - 12 i \omega M\right] 
    \phi_{RW} = \left[{\mu}^2 \left({\mu}^2 + 2\right) + {72 M^2 (r - 2 M) 
    \over r^2 \left({\mu}^2 r + 6 M\right)}\right] Z - 12 M {{\rm d} Z 
    \over {\rm d}r^*} \ ,
\end{equation}
where ${\mu}^2 = (l - 1) (l + 2)$ and $M$ is the mass of the star.

\vfill
\eject

\begin{tabular}{|c|c|c|c|c|c|c|c|c|}
\hline $Model \#$ & $\sigma_\n/m_\n$ & $\sigma_\p/m_\n$ & $\beta_\n$ & 
$\beta_\p$ & $\n_0({\rm fm}^{- 3})$ & $\p_0({\rm fm}^{- 3})$ & 
$M/M_{\odot}$ & $R({\rm km})$\\
\hline $one$ & $0.2$ & $0.5$ & $2.0$ & $2.0$ & $1.3$ & $0.52$ & $1.180$ & 
$8.657$\\
\hline $two$ & $0.2$ & $0.5$ & $2.5$ & $2.0$ & $1.3$ & $0.741$ & $1.355$ & 
$7.923$\\
\hline $three$ & $0.2$ & $0.5$ & $2.0$ & $2.0$ & $2.0$ & $0.8$ & $1.178$ & 
$7.651$\\
\hline $four$ & $0.2$ & $0.5$ & $2.5$ & $2.0$ & $2.0$ & $1.414$ & $1.30$ & 
$6.841$\\
\hline
\end{tabular}
\vskip 1.0cm
\noindent
{\bf Table I}.  Listed are four specific choices for the 
parameters of the relativistic two-polytrope master function given in 
Eq. (\ref{EOS}).  Also listed are the number densities, masses, and 
radii for the explicit solutions used in the quasi-normal mode 
analysis of the main text.

\vskip 1.0cm

\begin{tabular}{|c|c|c|}
\hline $Mode$ & $Re(\Delta \omega)/Re(\omega)\ (10^{-5})$ & $Im(\Delta 
\omega) /Im(\omega)\ (10^{-5})$ \\
\hline $1$ & $3.8$ & $2.0$ \\
\hline $2$ & $2.1$ & $2.4$ \\
\hline $3$ & $0.6$ & $5.4$ \\
\hline $4$ & $1.6$ & $4.9$ \\
\hline $5$ & $1.8$ & $9.8$ \\
\hline $6$ & $3.1$ & $12$ \\
\hline
\end{tabular}
\vskip 1.0cm
\noindent
{\bf Table II}.  Listed are the relative changes in the real and 
imaginary parts of $\omega$ for the first six w-modes when the step 
sizes are changed from $\Delta r = 10^{-5}$ to $\Delta r = 5 \times 
10^{-6}$. 

\vskip 1.5cm

\begin{tabular}{|c|c|c|c|c|}
\hline ${}$ & 
\multicolumn{2}{c|}{ $Model\ one$ } & 
\multicolumn{2}{c|}{ $Model\ two$ } \\
\hline
$Mode$ & ${\rm Re}(\omega M)$ & ${\rm Im}(\omega M)$
& ${\rm Re}(\omega M)$ & ${\rm Im}(\omega M)$ \\
\hline
${\rm w_{II1}}$ & $2.05\times 10^{-2}$ & $0.692$ & $0.151$
& $0.775$ \\
\hline 
${\rm w_{II2}}$ & $0.327$ & $0.380$ & $0.428$ & $0.397$ \\
\hline
${\rm w_1}$ & $0.497$ & $0.265$ & $0.510$ & $0.194$ \\
\hline
${\rm w_2}$ & $0.848$ & $0.367$ & $0.853$ & $0.316$ \\
\hline
${\rm w_3}$ & $1.183$ & $0.421$ & $1.189$ & $0.369$ \\
\hline
${\rm w_4}$ & $1.514$ & $0.460$ & $1.523$ & $0.406$ \\
\hline
\end{tabular}
\vskip 1.0cm
\noindent
{\bf Table III}. Listed are the frequencies, for $l = 2$, of the w-modes 
for models one and two of Table I. ${\rm w_{II1}}$ and ${\rm w_{II2}}$ 
are the strongly damped modes of Fig. 3 with ${\rm Im}(\omega M) > 
{\rm Re}(\omega M)$; ${\rm w_1}$, ${\rm w_2}$, etc., are likewise the 
first four modes of Fig. 3 (reading left to right) whose frequencies 
satisfy the opposite inequality ${\rm Im}(\omega M) < {\rm Re}(\omega M)$. 

\vskip 1.5 cm

\begin{tabular}{|c|c|c|c|c|}
\hline $Model \#$ & $Mode$ & $l=2$ & $l=3$ & $l=4$ \\
\hline
$one$ & ${\rm f}$ & $0.108$ & $0.142$ & $0.166$ \\
& ${\rm p_1}$ & $0.234$ & $0.281$ & $0.319$ \\
\hline
$two$ & ${\rm f_o}$ & $0.136$ & $0.180$ & $0.212$ \\
& ${\rm f_s}$ & $0.157$ & $0.192$ & $0.221$ \\
& ${\rm p_{1o}}$ & $0.306$ & $0.365$ & $0.414$ \\
& ${\rm p_{1s}}$ & $0.354$ & $0.417$ & $0.472$ \\
\hline
\end{tabular}
\vskip 1.0cm
\noindent
{\bf Table IV}. The real parts of the frequencies ($\omega M$) of the 
f- and ${\rm p_1}$-modes for $l = 2,3,4$ for models one and two of 
Table I. 

\vfill
\eject
\noindent

\vskip 0.5 cm
\noindent

\begin{figure}[t]
\centerline{\epsfig{file=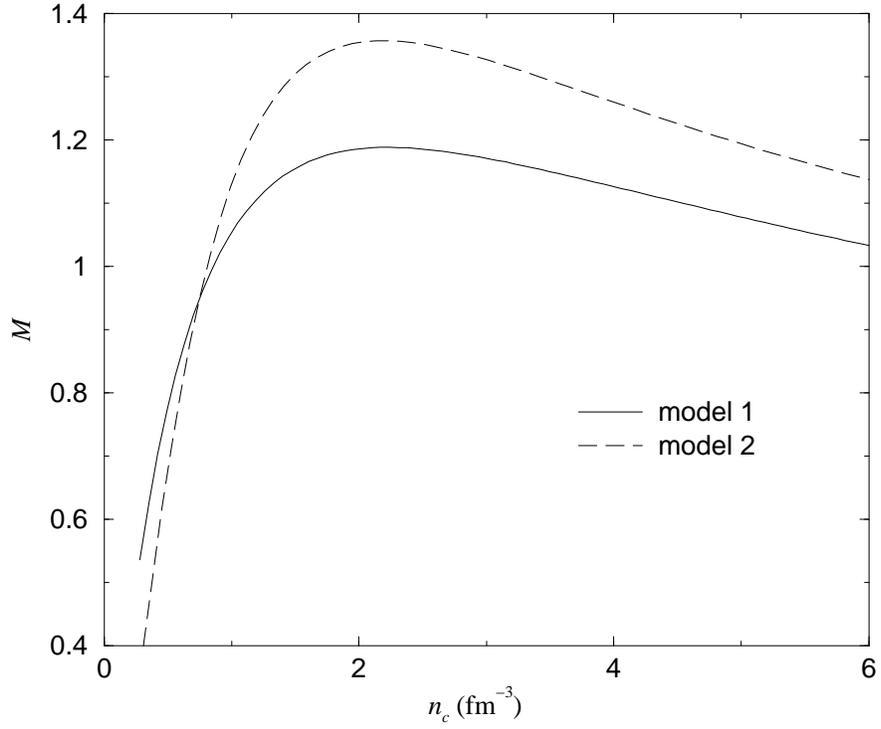,width=0.65\linewidth}}
\caption{\small 
$M$ vs $n_c = n_0 + p_0$ for the parameter values
given for models one and two in Table I. $M$ is given in units of
$M_{\odot}$.}
\end{figure}
\vskip 0.5 cm
\noindent

\begin{figure}[t]
\centerline{\epsfig{file=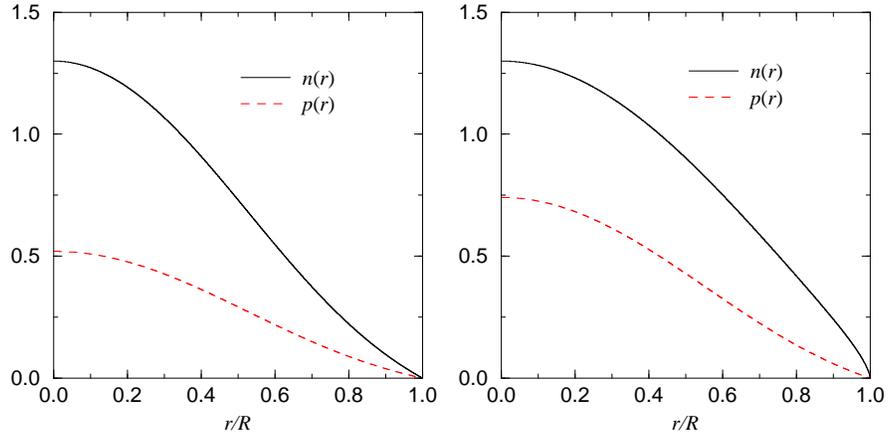,width=0.65\linewidth}}
\caption{\small
Number density profiles $n(r)$ and $p(r)$ for models 
one (left) and two (right) of Table I. $n$ and $p$ are given in 
units of ${\rm fm}^{-3}$.}
\end{figure}
\vskip 0.5 cm
\noindent

\begin{figure}[t]
\centerline{\epsfig{file=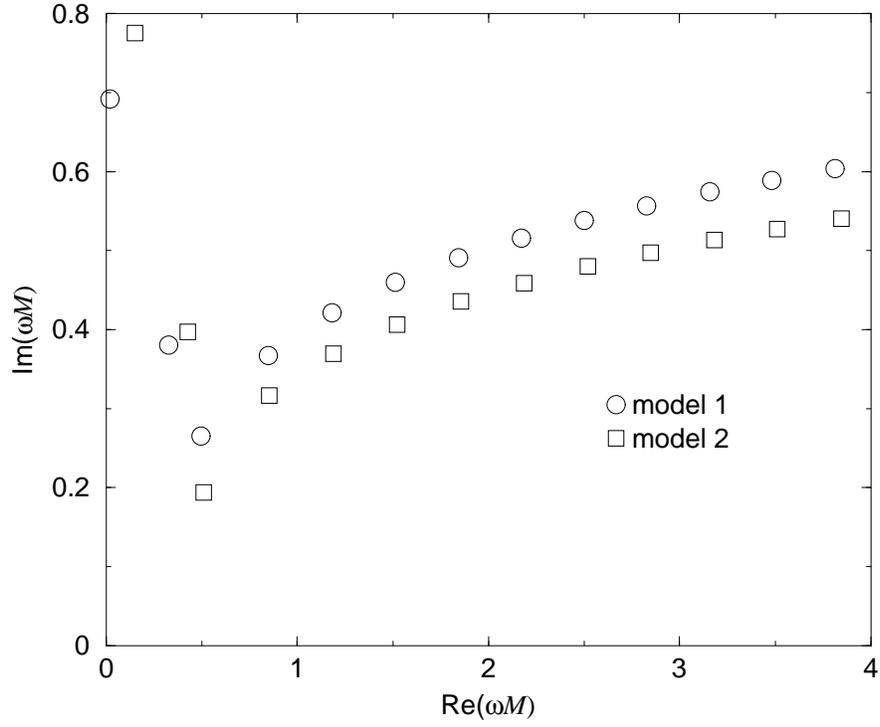,width=0.65\linewidth}}
\caption{\small
Im($\omega M$) vs Re($\omega M$) for the w-modes for 
models one and two described in Table I, setting $l = 2$.}
\end{figure}
\vskip 0.5 cm
\noindent

\begin{figure}[t]
\centerline{\epsfig{file=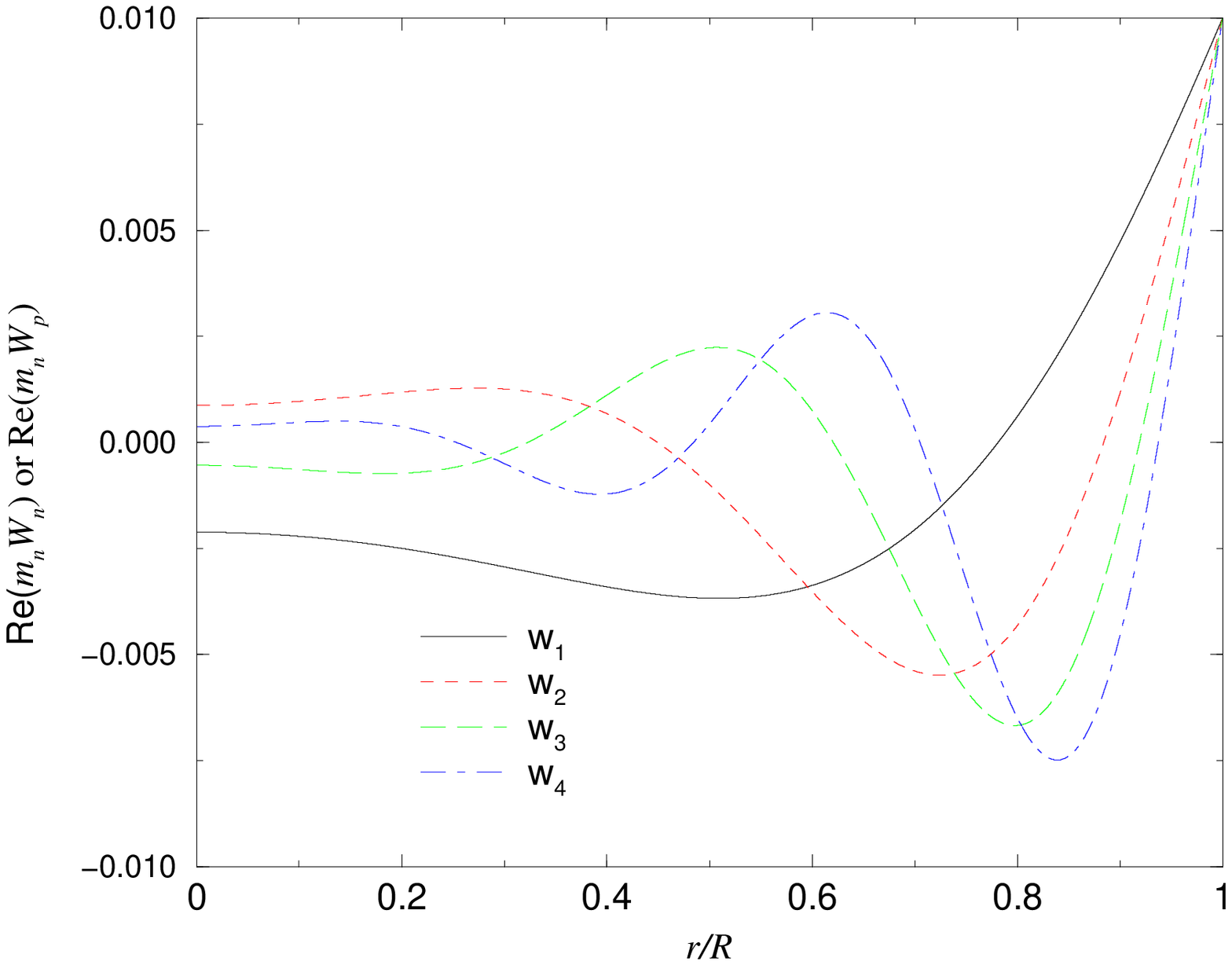,width=0.65\linewidth}}
\caption{\small
Re($m_nW_n$) or Re($m_nW_p$) (since they are indistinguishable) 
vs $r/R$ for the model one modes ${\rm w_1}$ through ${\rm w_4}$ listed 
in Table III.}
\end{figure}
\vskip 0.5 cm
\noindent 

\begin{figure}[t]
\centerline{\epsfig{file=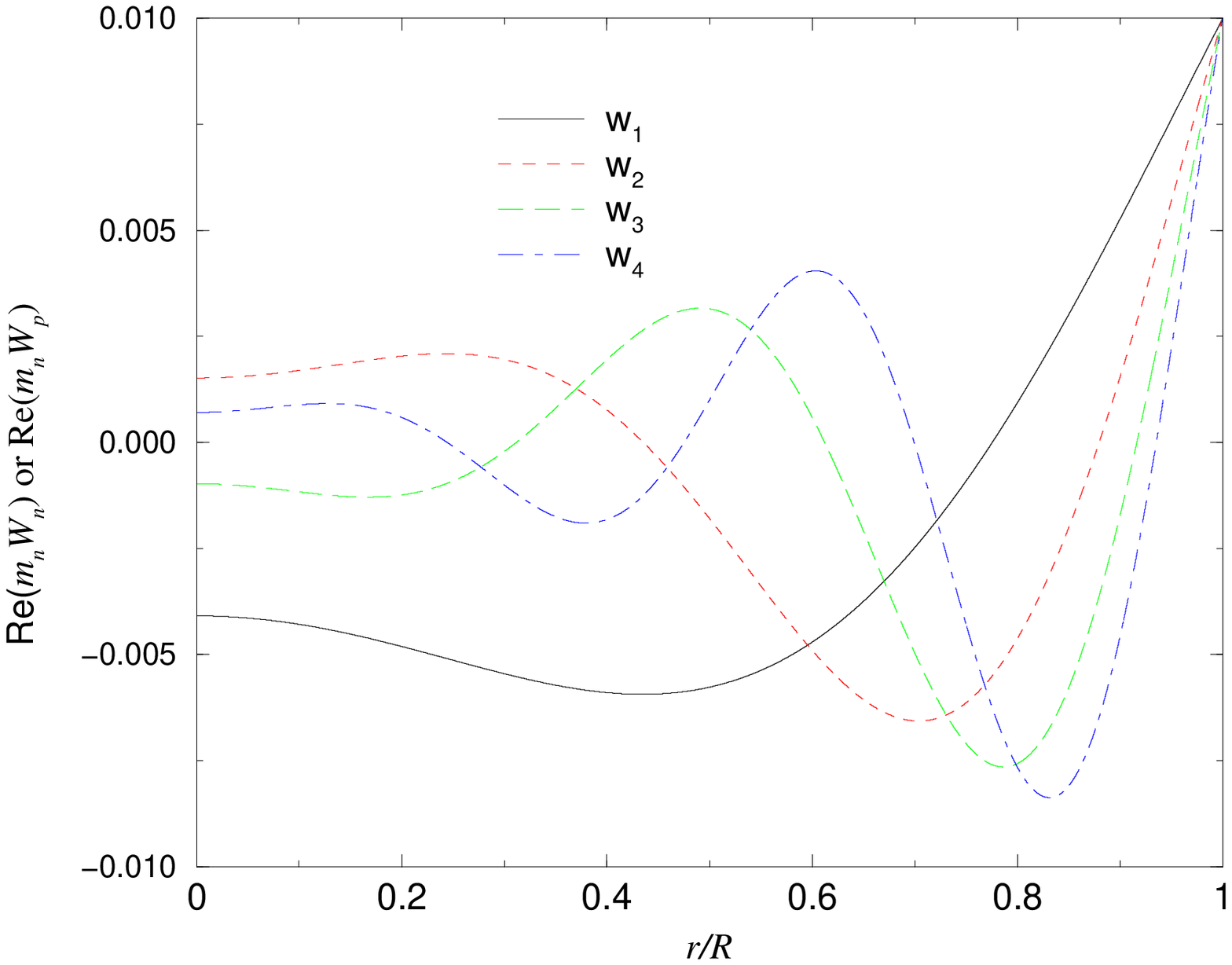,width=0.65\linewidth}}
\caption{\small
Re($m_nW_n$) or Re($m_nW_p$) (since they are indistinguishable) 
vs $r/R$ for the model two modes ${\rm w_1}$ through ${\rm w_4}$ listed 
in Table III.}
\end{figure}
\vskip 0.5 cm
\noindent

\begin{figure}[t]
\centerline{\epsfig{file=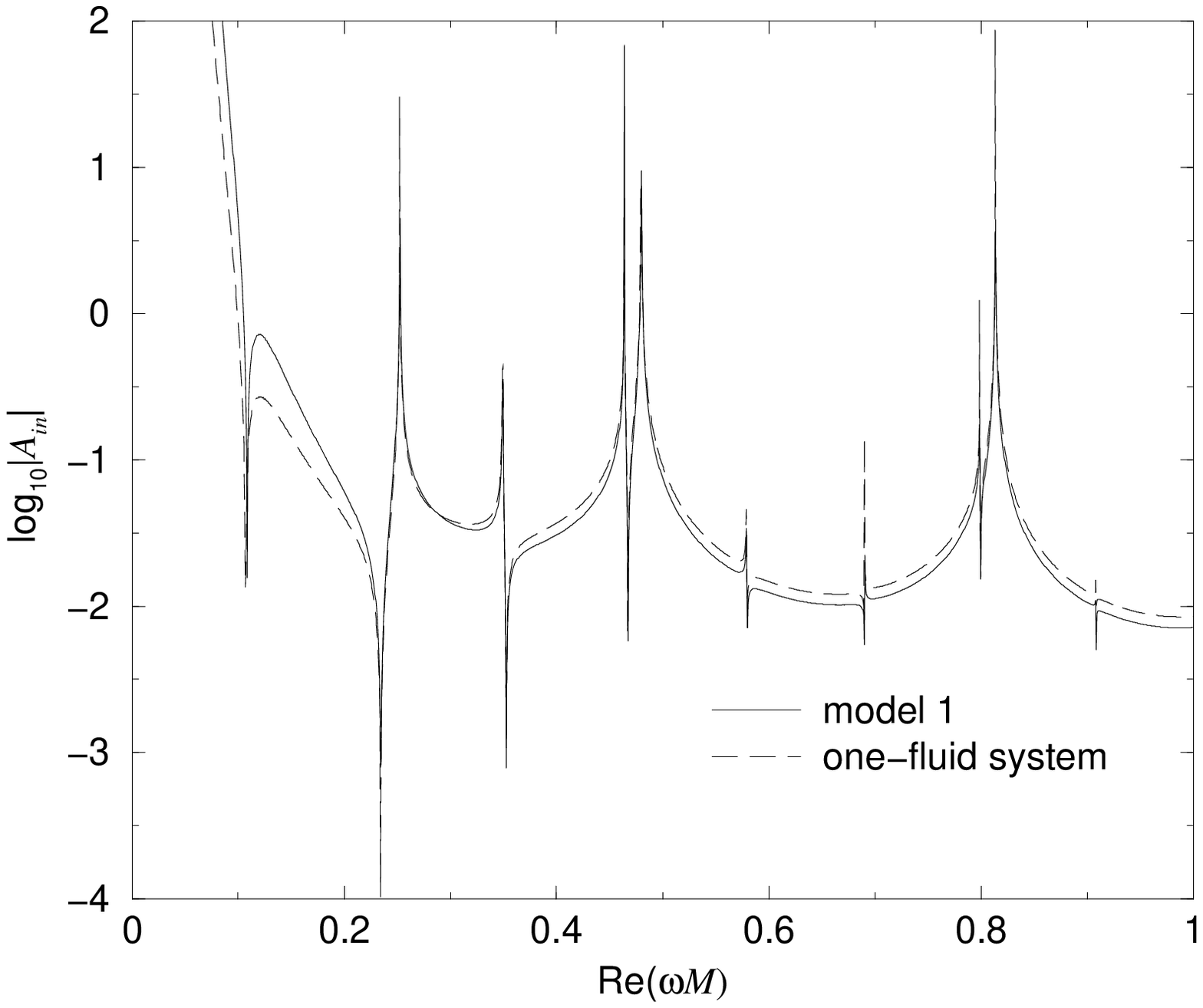,width=0.65\linewidth}}
\caption{\small
${\rm log}_{10}|A_{in}|$ vs ${\rm Re}(\omega M)$, with $l = 2$,
for model one of Table I, and for a one-fluid system composed
of neutrons behaving as a relativistic polytrope
(i.e. $\beta_n=\beta_p$, $\sigma_n=\sigma_p$, and with the same
values for $n_0$, $\beta_n$, and $\sigma_n$ as model one).}
\end{figure}
\vskip 0.5 cm
\noindent

\begin{figure}[t]
\centerline{\epsfig{file=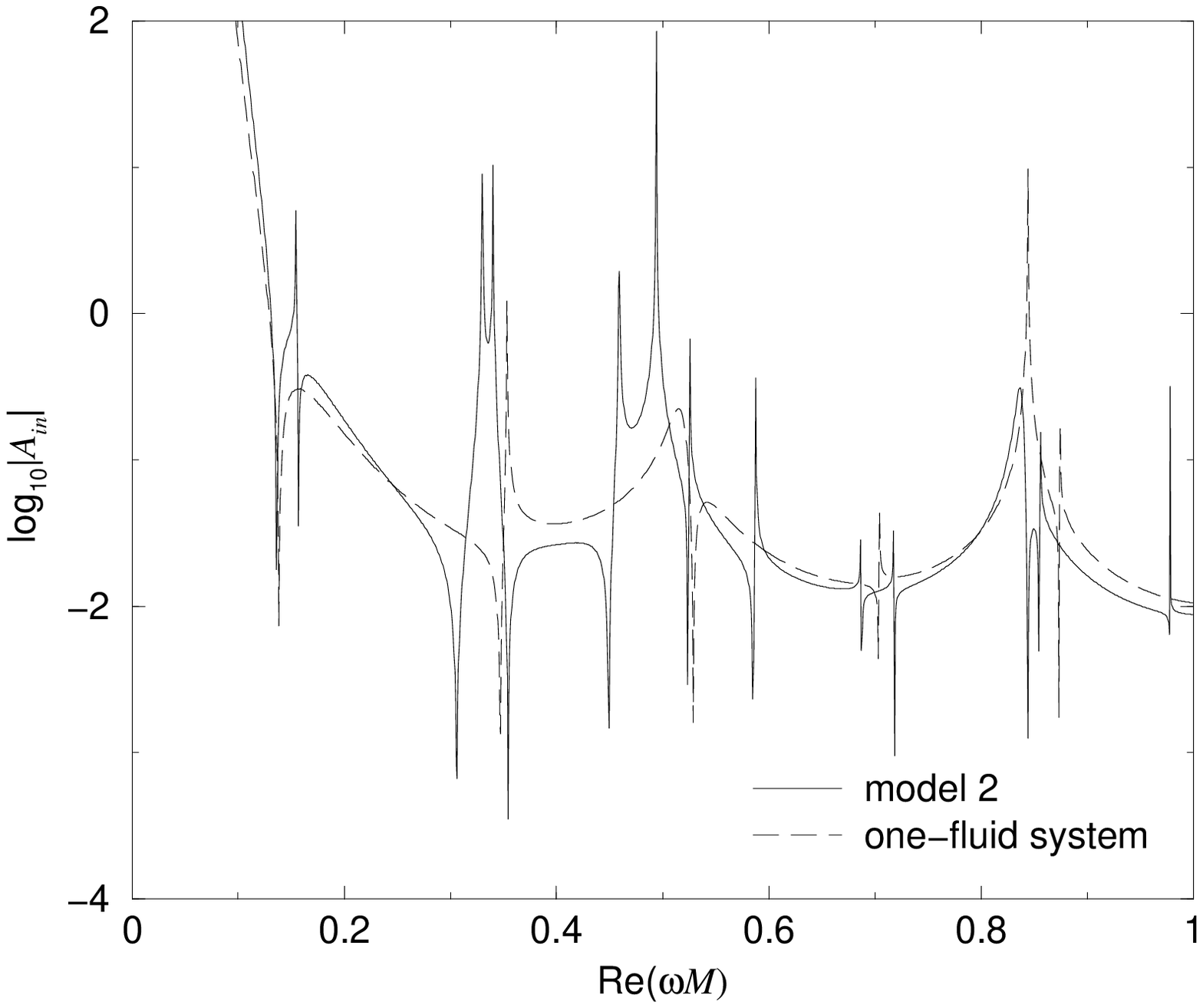,width=0.65\linewidth}}
\caption{\small
${\rm log}_{10}|A_{in}|$ vs ${\rm Re}(\omega M)$, with $l = 2$,
for model two of Table I, and for a one-fluid system composed
of neutrons behaving as a relativistic polytrope
(i.e. $\beta_n=\beta_p$, $\sigma_n=\sigma_p$, and with the same
values for $n_0$, $\beta_n$, and $\sigma_n$ as model two).}
\end{figure}
\vskip 0.5 cm
\noindent

\begin{figure}[t]
\centerline{\epsfig{file=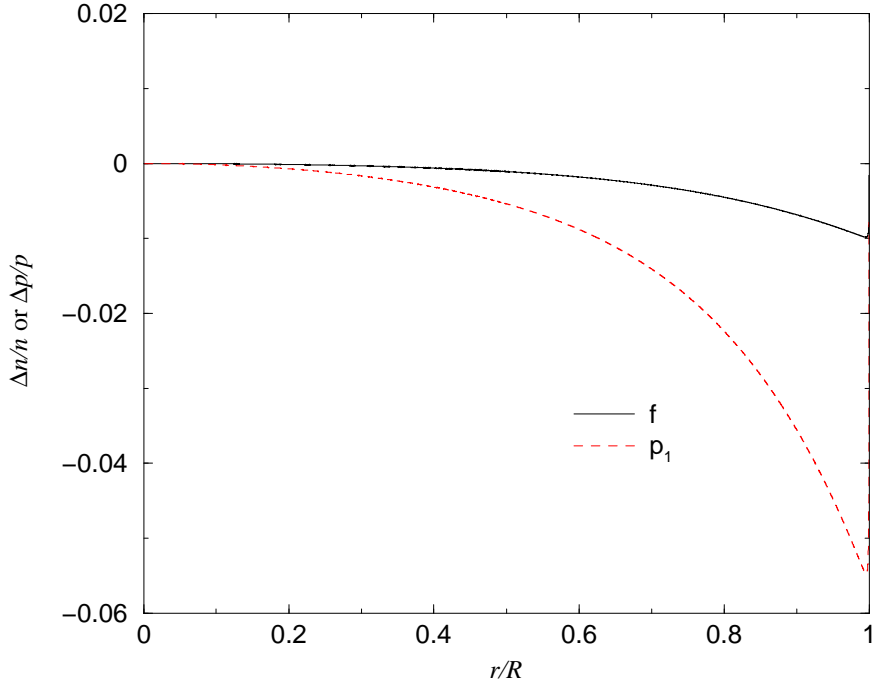,width=0.65\linewidth}}
\caption{\small
$\Delta n/n$ or $\Delta p/p$ (since they are indistinguishable)
vs $r/R$, with $l = 2$, for the model one f- and ${\rm p_1}$- modes 
listed in Table IV.}
\end{figure}
\vskip 0.5 cm
\noindent

\begin{figure}[t]
\centerline{\epsfig{file=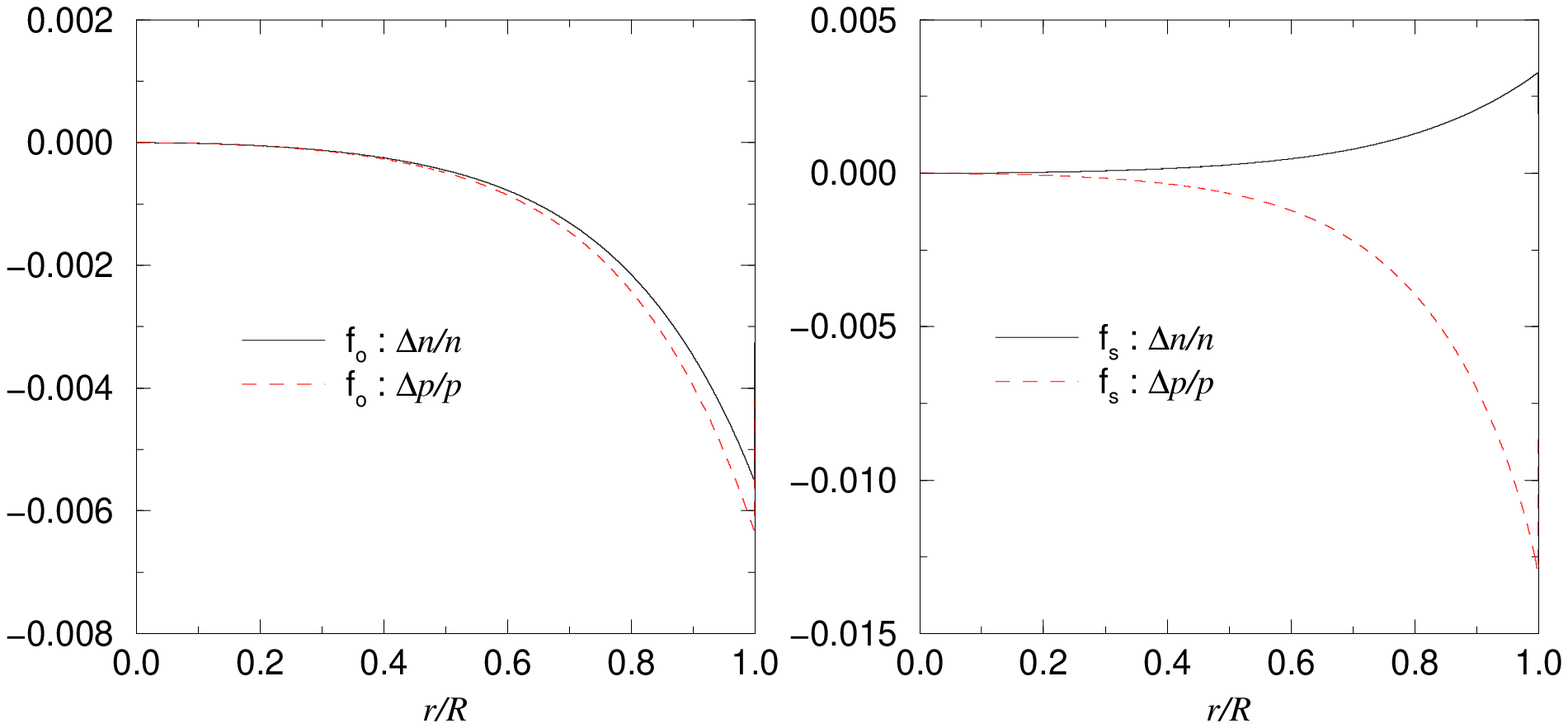,width=0.65\linewidth}}
\caption{\small
$\Delta n/n$ and $\Delta p/p$ vs $r/R$, with $l = 2$, for the 
model two ${\rm f_o}$- and ${\rm f_s}$-modes listed in Table IV.}
\end{figure}
\vskip 0.5 cm
\noindent

\begin{figure}[t]
\centerline{\epsfig{file=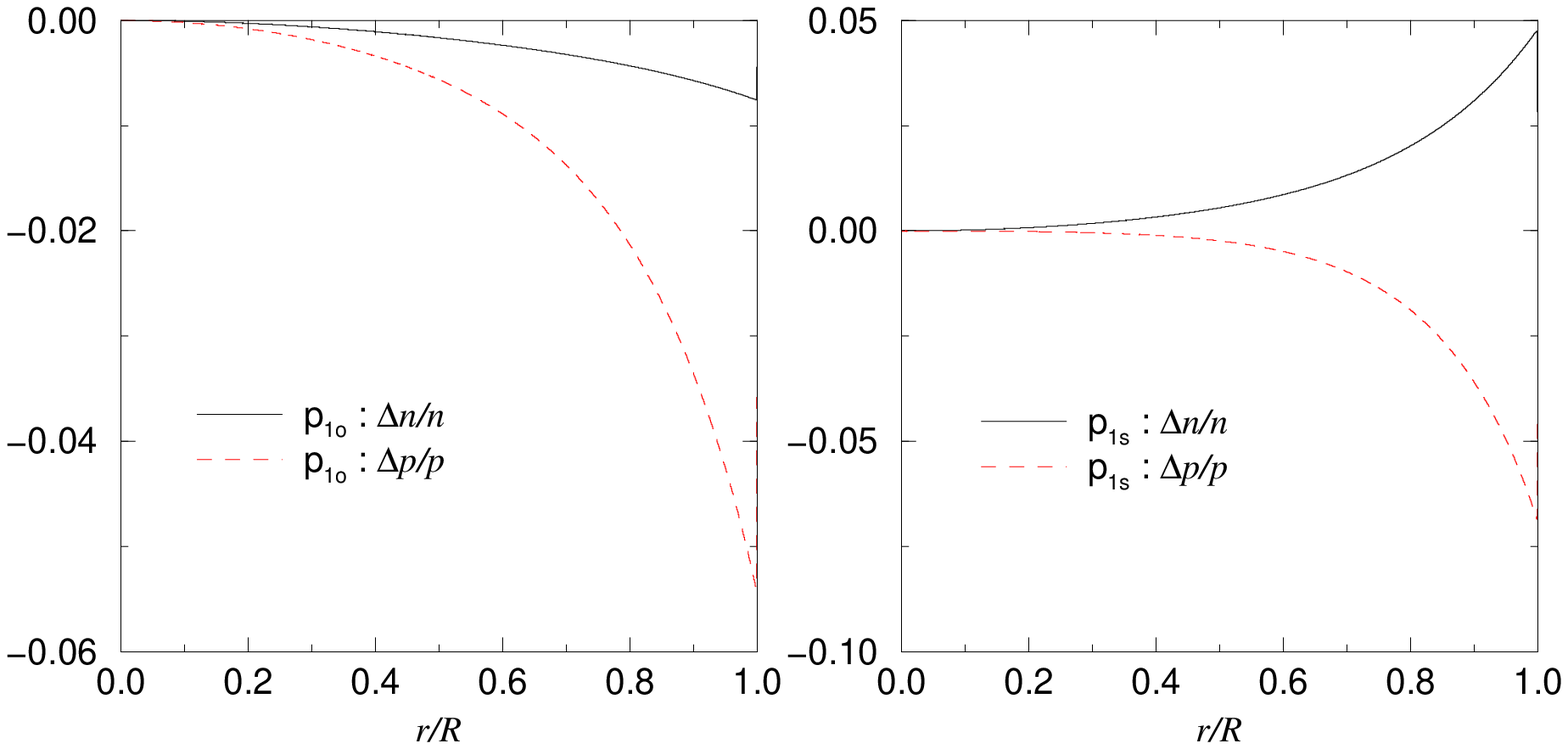,width=0.65\linewidth}}
\caption{\small
$\Delta n/n$ and $\Delta p/p$ vs $r/R$, with $l = 2$, for the 
model two ${\rm p_{1o}}$- and ${\rm p_{1s}}$-modes listed in Table IV.}
\end{figure}
\vskip 0.5 cm
\noindent

\begin{figure}[t]
\centerline{\epsfig{file=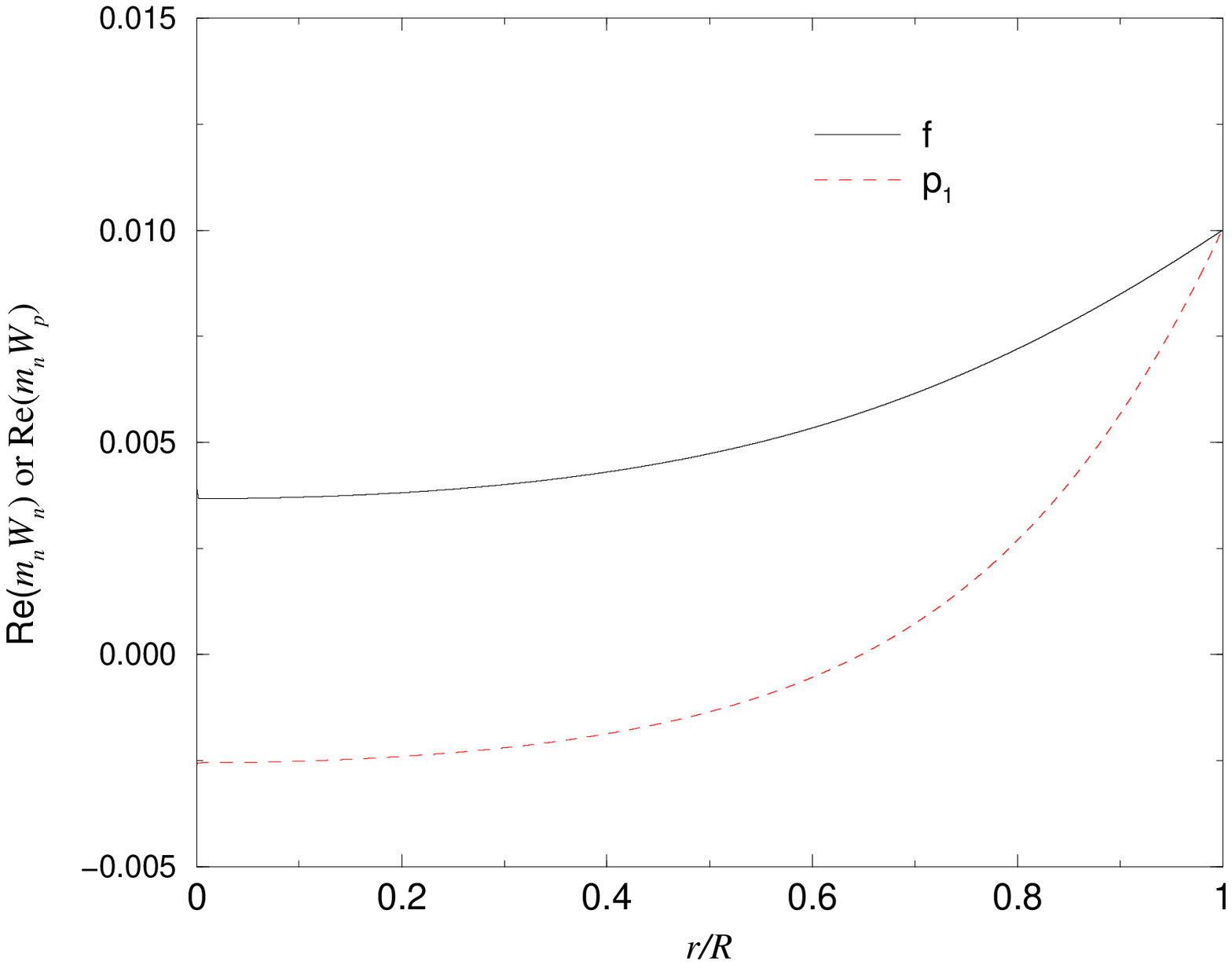,width=0.65\linewidth}}
\caption{\small
Re($m_nW_n$) or Re($m_nW_p$) (since they are indistinguishable)
vs $r/R$, with $l = 2$, for the model one f- and ${\rm p_1}$-modes  
listed in Table IV.}
\end{figure}
\vskip 0.5 cm
\noindent

\begin{figure}[t]
\centerline{\epsfig{file=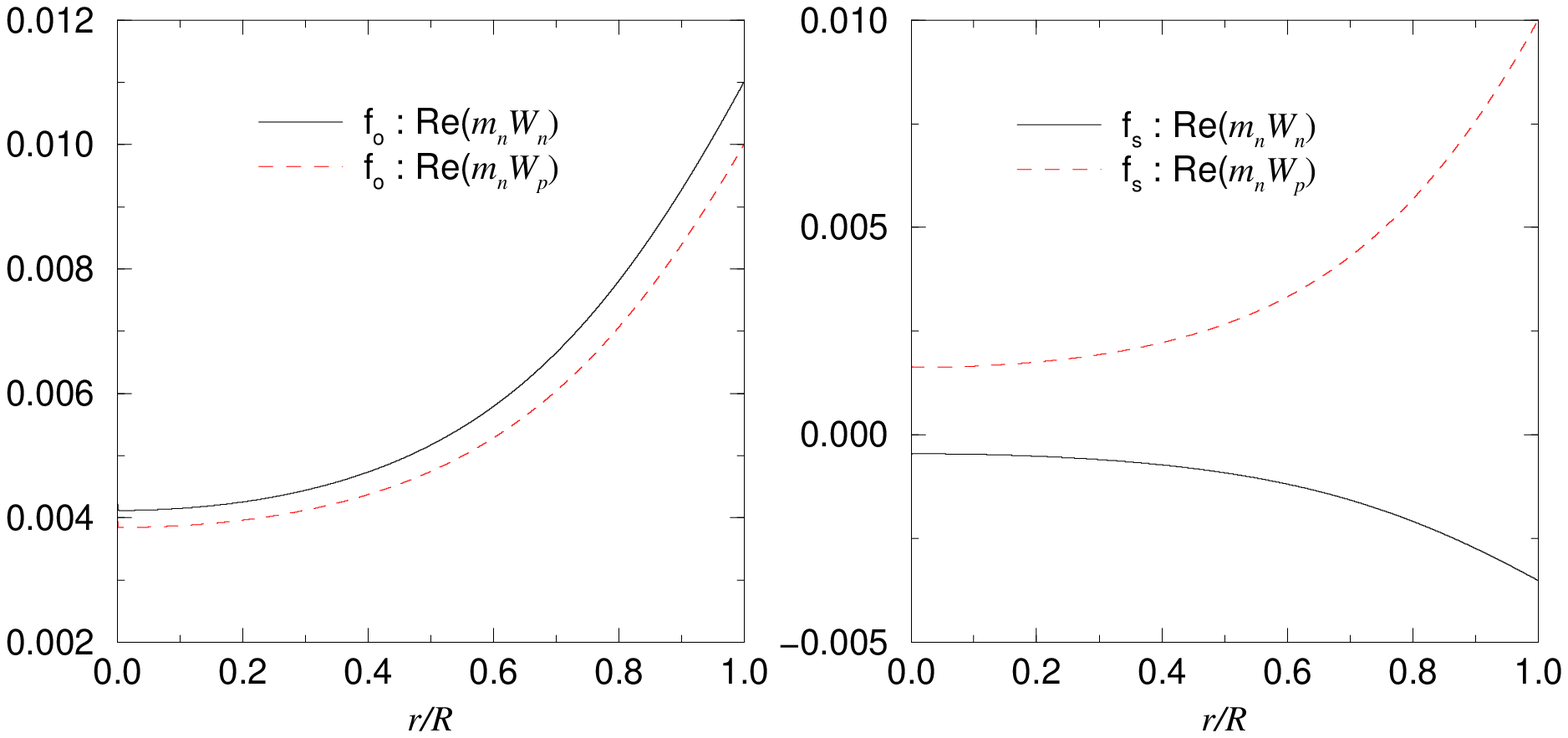,width=0.65\linewidth}}
\caption{\small
Re($m_nW_n$) and Re($m_nW_p$) vs $r/R$, with $l = 2$, for the 
model two ${\rm f_o}$- and ${\rm f_s}$-modes listed in Table IV.}
\end{figure}
\vskip 0.5 cm
\noindent

\begin{figure}[t]
\centerline{\epsfig{file=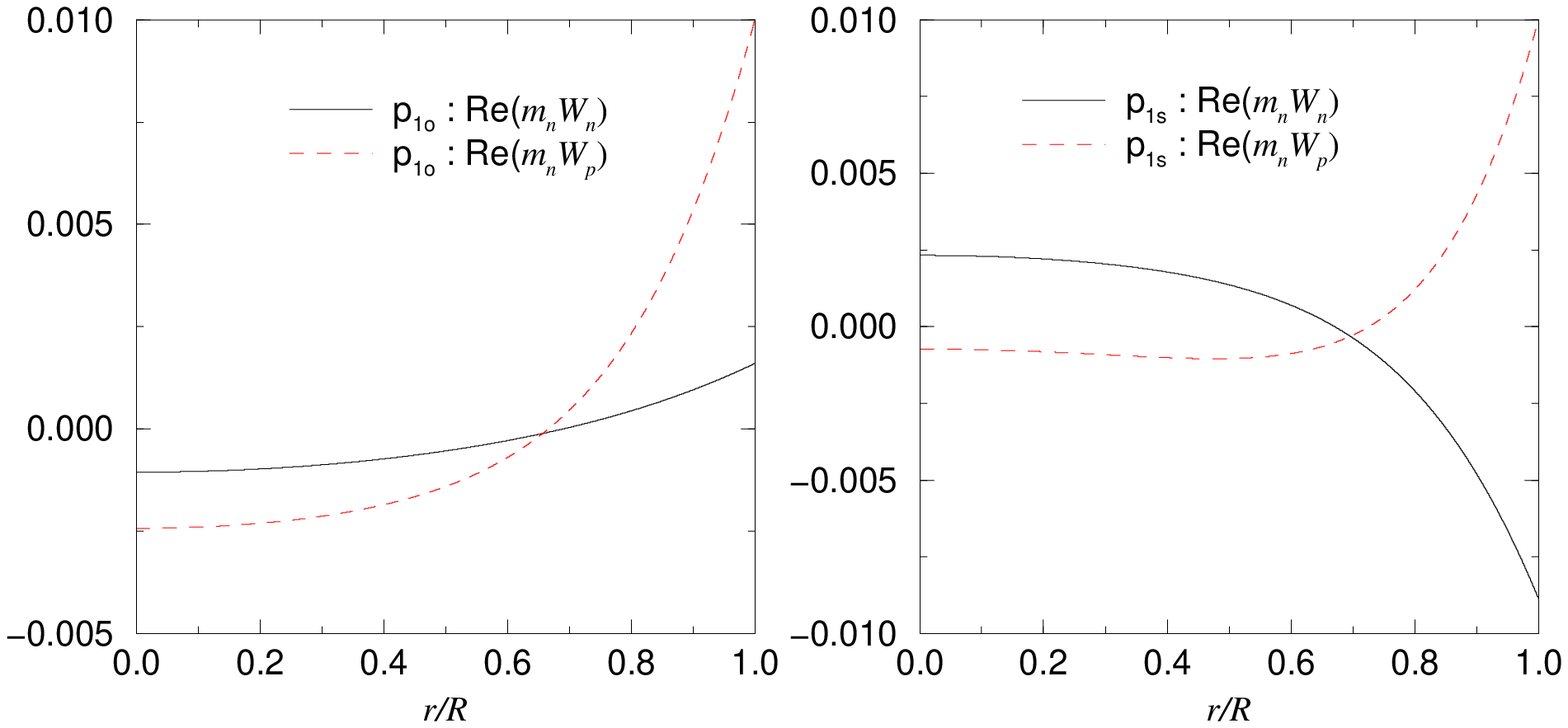,width=0.65\linewidth}}
\caption{\small
Re($m_nW_n$) and Re($m_nW_p$) vs $r/R$, with $l = 2$, for the 
model two ${\rm p_{1o}}$- and ${\rm p_{1s}}$-modes listed in Table IV.}
\end{figure}
\vskip 0.5 cm
\noindent

\begin{figure}[t]
\centerline{\epsfig{file=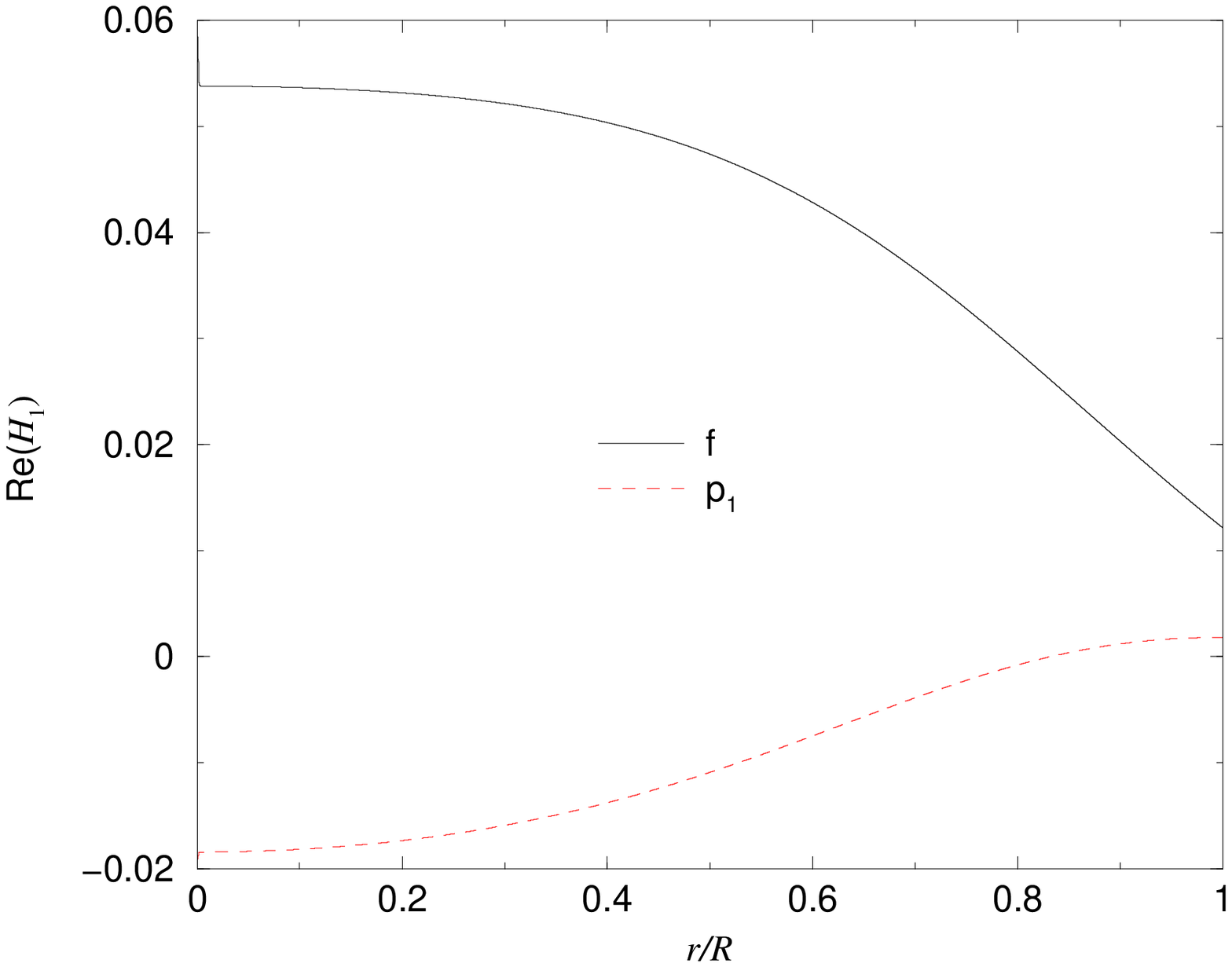,width=0.65\linewidth}}
\caption{\small
Re($H_1$) vs $r/R$, with $l = 2$, for the model one f- and 
${\rm p_1}$-modes listed in Table IV.}
\end{figure}
\vskip 0.5 cm
\noindent

\begin{figure}[t]
\centerline{\epsfig{file=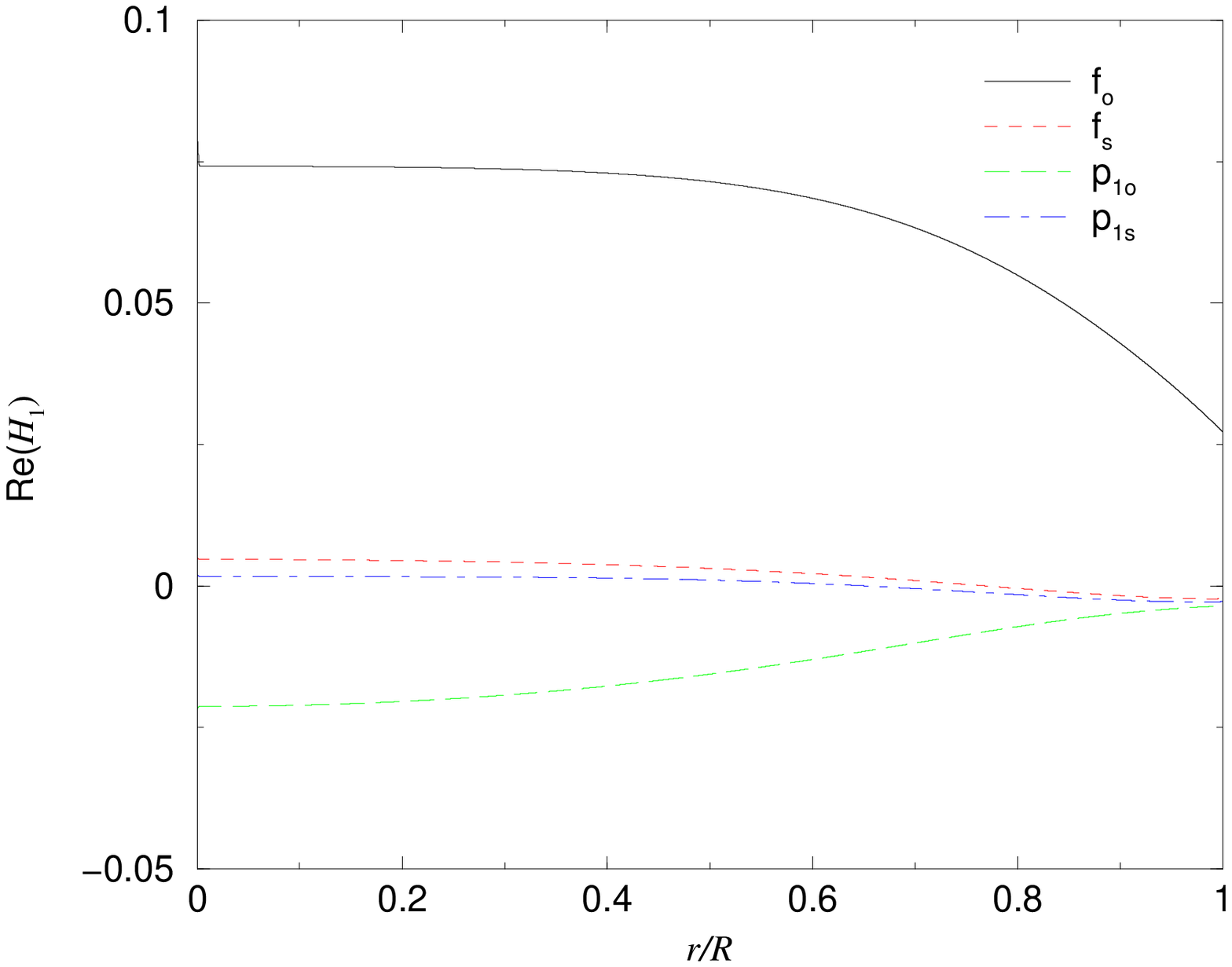,width=0.65\linewidth}}
\caption{\small
Re($H_1$) vs $r/R$, with $l = 2$, for the model two 
${\rm f_o}$-, ${\rm f_s}$-, ${\rm p_{1o}}$-, and ${\rm p_{1s}}$-modes  
listed in Table IV.}
\end{figure}
\vskip 0.5 cm
\noindent

\begin{figure}[t]
\centerline{\epsfig{file=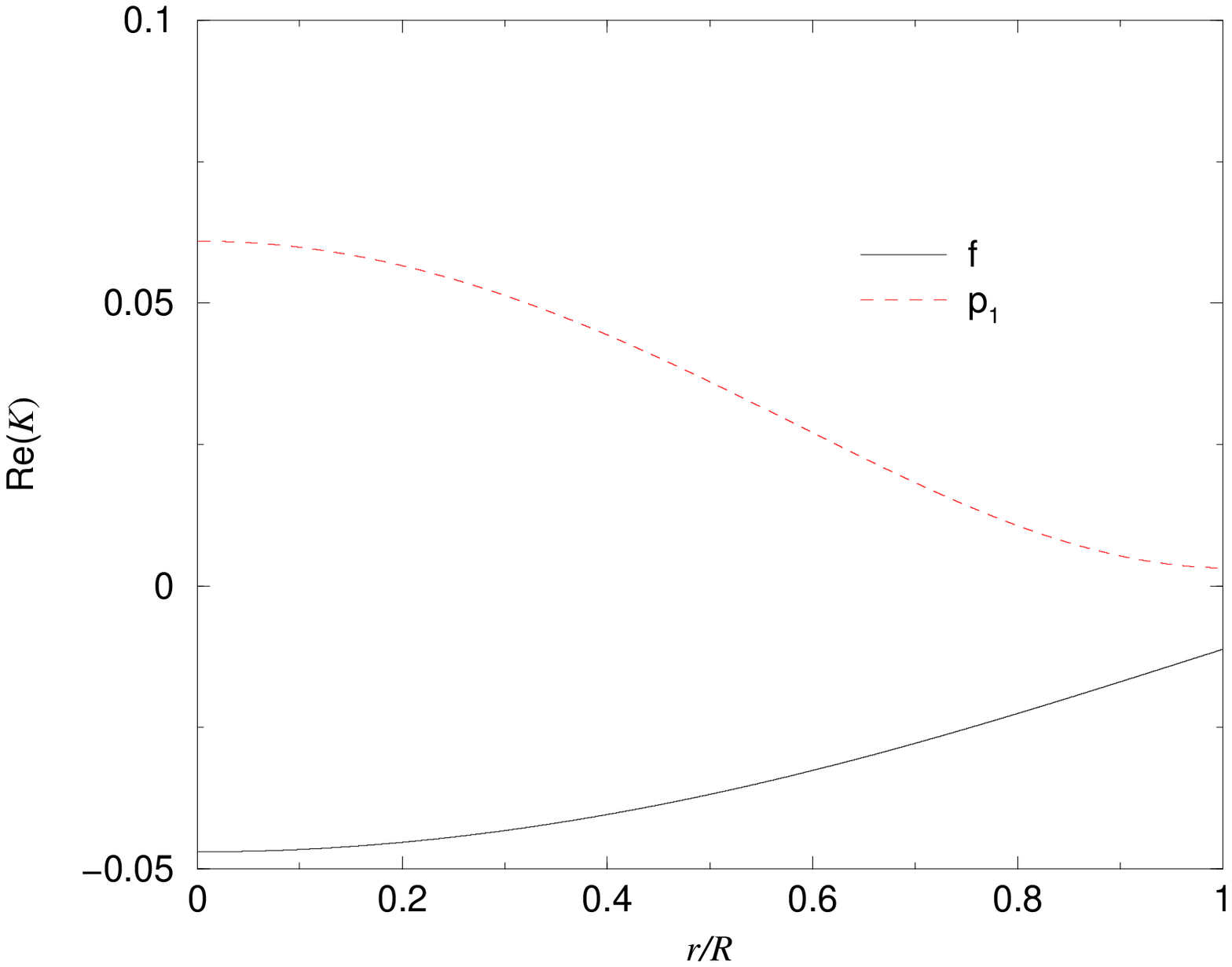,width=0.65\linewidth}}
\caption{\small
Re($K$) vs $r/R$, with $l = 2$, for the model one f- and 
${\rm p_1}$-modes listed in Table IV.}
\end{figure}
\vskip 0.5 cm
\noindent

\begin{figure}[t]
\centerline{\epsfig{file=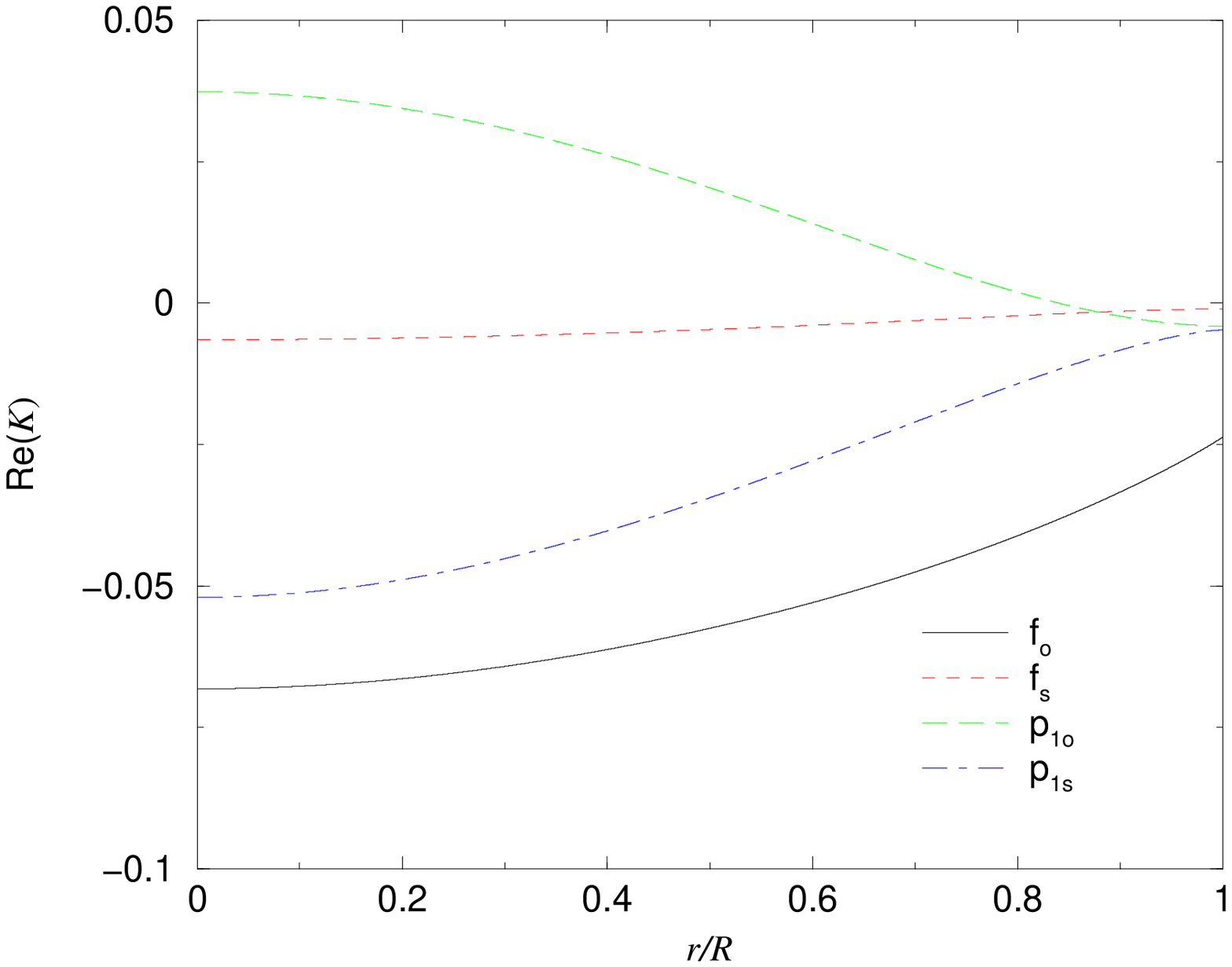,width=0.65\linewidth}}
\caption{\small
Re($K$) vs $r/R$, with $l = 2$, for the model two 
${\rm f_o}$-, ${\rm f_s}$-, ${\rm p_{1o}}$-, and ${\rm p_{1s}}$-modes  
listed in Table IV.}
\end{figure}

\end{document}